\documentclass[preprint2]{aastex}

\slugcomment{To appear in ApJ}

\shorttitle{The outskirts of NGC1399}
\shortauthors{Iodice et al.}

\begin{document}


\title{The Fornax Deep Survey with VST. I. \\ The extended and diffuse stellar halo of NGC~1399 out to 192 kpc.}


%\author{E. Iodice\altaffilmark}
%\email{iodice@na.astro.it}

\author{E. Iodice\altaffilmark{1}, M. Capaccioli\altaffilmark{2}, A. Grado\altaffilmark{1}, L. Limatola\altaffilmark{1}, M. Spavone\altaffilmark{1}, N.R. Napolitano\altaffilmark{1}, M. Paolillo\altaffilmark{2,8,9}, R. F. Peletier\altaffilmark{3}, M. Cantiello\altaffilmark{4}, T. Lisker\altaffilmark{5}, C. Wittmann\altaffilmark{5}, A. Venhola\altaffilmark{3,6}, M. Hilker\altaffilmark{7}, R. D'Abrusco\altaffilmark{2}, V. Pota\altaffilmark{1}, P. Schipani\altaffilmark{1}}

%\affil{INAF - Astronomical Observatory of Capodimonte, via Moiariello 16, Naples, I-80131, Italy}
%\and
%\author{M. Paolillo}
%\affil{University of Naples Federico II, C.U. Monte SantÕAngelo, Via Cinthia, 80126, Naples, Italy}
%\and
%\author{ R. Peletier}
%\affil{Kapteyn Astronomical Institute, University of Groningen, PO Box 72, 9700 AB Groningen, The Netherland}
%\and
%\author{M. Cantiello}
%\affil{INAF-Astronomical Observatory of Teramo, Via Maggini, 64100, Teramo, Italy}
%\and
%\author{T. Lisker}
%\and
%\author{C. Wittmann}
%\affil{...}

%\affil{Space Telescope Science Institute, Baltimore, MD 21218}


\altaffiltext{1}{INAF - Astronomical Observatory of Capodimonte, via Moiariello 16, Naples, I-80131, Italy}
\altaffiltext{2}{Dip.di Fisica ÒEttore PanciniÓ, University of Naples ''Federico II'', C.U. Monte SantÕAngelo, Via Cinthia, 80126, Naples, Italy}
\altaffiltext{3}{Kapteyn Astronomical Institute, University of Groningen, PO Box 72, 9700 AV Groningen, The Netherlands}
\altaffiltext{4}{INAF - Astronomical Observatory of Teramo, Via Maggini, 64100, Teramo, Italy}
\altaffiltext{5}{Zentrum fuer Astronomie der Universitaet Heidelberg, Germany}
\altaffiltext{6}{Division of Astronomy, Department of Physics, University of Oulu, Oulu, Finland}
\altaffiltext{7}{European Southern Observatory, Karl-Schwarzschild-Strasse 2, D-85748 Garching bei MŸnchen, Germany}
\altaffiltext{8}{INFN Sezione di Napoli, Via Cintia, I-80126 Napoli, Italy}
\altaffiltext{9}{Agenzia Spaziale Italiana - Science Data Center, Via del Politecnico snc, 00133 Roma, Italy}


\begin{abstract}
We have started a new deep, multi-imaging survey of the Fornax cluster, dubbed Fornax Deep Survey (FDS),   at the VLT Survey Telescope. In this paper we present the deep photometry inside two square degrees around the bright  galaxy NGC~1399 in the core of the cluster. 
We found that the core of the Fornax cluster is characterised by a very extended and diffuse envelope surrounding the luminous galaxy NGC~1399: we map the surface brightness out to 33~arcmin ($\sim 192$~kpc) from the galaxy center and down to $\mu_g \sim 31$~mag arcsec$^{-2}$ in the $g$ band.
The deep photometry allows us to detect a faint stellar bridge  in the intracluster region on the west side of NGC~1399 and towards NGC~1387. By analyzing the integrated colors of this feature, we argue that it could be due to the ongoing interaction between the two galaxies, where the outer envelope of NGC~1387 on its east side is stripped away.
By fitting the light profile,  we found that it exists a physical break radius in the total light distribution at $R=10$~arcmin ($\sim58$~kpc) that sets the transition region between the bright central galaxy and the outer exponential stellar halo.  
We discuss the main implications of this work on the build-up of the stellar halo at the center of the Fornax cluster. By comparing with the numerical simulations of the stellar halo formation for the most massive BCGs (i.e. $13 < \log M_{200}/M_{\odot} < 14$), we find that the observed stellar halo mass fraction is consistent with a halo formed through the multiple accretion of progenitors with  stellar mass in the range $10^{8} - 10^{11}$~M$_{\odot}$. This might suggest that the halo of NGC~1399 has also gone through a major merging event. The absence of a significant number of luminous stellar streams and tidal tails out to 192~kpc suggests that the epoch of this strong interaction goes back to an early formation epoch. Therefore, differently from the Virgo cluster, the extended stellar halo around NGC~1399 is characterised by a more diffuse  and  well-mixed  component, including the ICL.

\end{abstract}





\section{Introduction}\label{intro}

%- Review on the halos of galaxies

Deep and large-scale multi-band imaging are crucial to study the galaxy outskirts, out to hundreds of kiloparsecs, where the imprints of the mass assembly reside: these are the regions of the stellar halos. 
Under the hierarchical accretion scenario,  galaxies at the center of the clusters continue to undergo active mass assembly and their halos are still growing in the present epoch: stellar halos and the diffuse intracluster light (ICL) results from the stripping of stars off of the cluster members by minor mergers and dynamical harassment \citep{Delucia2007,Puchwein2010,Cui2014}. 

Observations show that stellar halos can be made of multiple stellar components, can have complex kinematics and host substructures, in the form of shells and tidal tails, which indicate gravitational interactions, like merging and/or accretion, in the formation history of a galaxy \citep[see][and references therein]{Bender2015,Longobardi2015}. The light distribution of the stellar halos can appear as a different component at large radii with respect to the inner light profile, showing an exponential decrease \citep{Seigar2007,Donzelli2011}, or, alternatively, they result as shallower outer region of the whole light profile, which can be well fitted by a single Sersic law with a high value of the $n$ exponent, $n \sim 7-8$, \citep{Bender2015}.
The relics of the interactions, such as tails, shells or streams, are also very faint, with a typical surface brightness  below $\mu_V \sim 27$~mag~arcsec$^{-2}$. Therefore, their detection requires very deep imaging and a careful data reduction, able to perform the sky subtraction with an accuracy of a few percent.

In the recent years, a big effort was made to develop deep photometric surveys aimed to study galaxy structures up to the faintest levels of surface brightness, which reveal such a kind of low luminosity structures and trace the stellar halos at very large distances from the galaxy center 
\citep{Mihos2005,Jan2010,MarDel2010,Roediger2011,Ferrarese2012,Duc2015,Dokkum2014,Munoz2015,Trujillo2015}.
Recently, deep images of the Virgo cluster have revealed several faint ($\mu_V = 26-29$~mag~arcsec$^{-2}$)  streams of ICL among the galaxy members and made it possible  to map the stellar halos of the bright cluster galaxies (BCGs) at very large distances from the center \citep{Jan2010}. In particular, the halo of M87 is traced out to about 150~kpc \citep{Mihos2015}. 
A steep color gradient towards bluer colors is measured in the halo regions, suggesting a significant contribution of metal-poor stars in its outskirts. 
In this context, the VEGAS survey, an ongoing multi-band survey of early-type galaxies in the Southern hemisphere \citep{Cap2015}, is producing competitive results. In particular, for the cD galaxy NGC~4472 (in the Virgo cluster) genuine new results concern the detection of an intracluster light tail between $5R_e\le R \le10 R_e$ , in the range of surface brightness of 26.5 - 27.6 mag/ arcsec$^2$ in the $g$ band. 

%-->discrete tracers 
For galaxies in the local universe ($\le 30$~Mpc), {  the properties of the stellar halos can also be  addressed  using discrete tracers} such as globular clusters (GCs) and planetary nebulae (PNe). Observations show that the red GCs populations have number density profile consistent with that of the stars of the parent galaxy, while the blue GCs are spatially more extended in the intra-cluster space and they trace the metal-poor component of the halo \citep{Forbes2012,Pota2013,Durrell2014,Hilker2015,Cantiello2015}.  In the Hydra cluster, the location of blue GCs coincides with a group of dwarf galaxies under disruption, suggesting a young stellar halo that is still forming \citep{Arnaboldi2012,Hilker2015}. In this case, this interpretation is further supported by kinematic measurements. In M87, the kinematics of PNe, their $\alpha$-parameter and the shape of the luminosity function, allow to separate the halo component from the ICL population \citep{Longobardi2015a}.
As stressed by \citet{Bender2015}, deep photometry alone is not enough to recognise the structure of the stellar halo, and kinematics measurement are necessary when possible.  In particular, the possibility to infer the variation of the orbital distribution of stars, PNe and GCs \citep[e.g.][]{Gerhard2001,Romanowsky2009,Napolitano2009,Napolitano2011,Napolitano2014,Pota2015} in the galaxy haloes is a strong observational test for galaxy formation scenarios that can be compared with predictions of hydrodynamical simulations \citep[e.g.][]{Wu2014}.

From the theoretical side, semi-analytic models combined with cosmological N-body simulations have become very sophisticated, with detailed predictions about the structure of stellar halos, the amount of substructure, the mass profiles, metallicity gradients, etc. \citep[e.g.]{Cooper2013,Pillepich2014}, in various types of environment. Recent studies have demonstrated that the overall structure of stellar halos, their stellar populations and the properties of their dynamical substructure directly probe two fundamental aspects of galaxy formation in the $\Lambda$CDM model: the hierarchical assembly of massive galaxies and their dark matter halos, and the balance between in situ star formation and accretion of stars through mergers \citep{Cooper2013,Cooper2015a,Cooper2015b,Pillepich2015}. 

%- FDS
The rich environments, clusters and groups of galaxies, are the ideal laboratory to probe the physical processes  leading to the formation of the stellar halos and the ICL. In this framework, we are performing  a deep multiband ($u$,$g$,$r$ and $i$) survey of the nearby Fornax cluster, at the VLT Survey Telescope (VST), situated at the Paranal observatory of the European Southern Observatory.
The {\it Fornax Deep Survey (FDS) with VST} aims to cover the whole Fornax cluster out to the virial radius \citep{Drinkwater2001}, with an area of about $26$~square degrees around the central galaxy NGC~1399,  including the region where NGC~1316 is located.  
The FDS will provide an unprecedented view of the structures of the cluster members, ranging from giant early-type galaxies to small spheroidal galaxies. The multi-band deep images will allow us to map the light distribution and colors of cluster galaxies out to 8-10 effective radii, in order to study the inner disks and bars,  the faint stellar halo, including the diffuse light component, and the tidal debris  as signatures of recent cannibalism events. Moreover, FDS will also give a full characterisation of small stellar systems around galaxies, as globular clusters (GCs) and ultra-compact dwarf galaxies (UCDs).

A detailed description of the FDS observations, data reduction and analyzis, and scientific goals is the subject of a forthcoming paper. In this work, it's urgent to present the first results of the FDS, focused on the bright cluster member NGC~1399, which is the cD galaxy in the core of Fornax cluster. This galaxy has been studied in detail in a wide wavelength range, from radio to X-ray. 
NGC~1399 is characterised by a large number of compact stellar systems (GCs and UCDs), for which several studies have provided the number density distribution and stellar population estimates \citep{Dirsch2003,Mieske2004,Bassino2006,Schuberth2010,Puzia2014,Voggel2015}. 

The photometry and kinematics of the inner and brighter regions of the galaxy resemble those for a typical elliptical galaxy \citep{Saglia2000}. 
The 2-dimensional light distribution of the central brightest regions of the galaxy appear very regular and round in shape \citep{Ferguson1989}. 
The most extended light profiles where published by  \citet{Schombert86} in the V band and  those by \citet{Caon1994} in the B band, mapping the surface brightness out to $R\sim40$~arcmin ($\sim 230$~kpc) and $R\sim 14$~arcmin ($\sim 81$~kpc), respectively. 
The most extended kinematics of stars (out to $\sim 90$~arcsec) was derived by \citet{Saglia2000}: it shows a small rotation ($\sim30$ km/s) and a high stellar velocity dispersion in the center ($\sim370$~km/s), which decreases at larger radii and remains constant at 250-270 km/s.
At larger radii, out to $\sim 13$~arcmin, the kinematics is traced by GCs and PNe \citep{Schuberth2010,McNeil2010}.
The kinematics of red and blue GCs are distinct, with red GCs having velocity dispersion consistent with that of stars, while blue GCs are more erratic, showing an higher velocity dispersion, $300 \le \sigma \le 400$~km/s, \citep{Schuberth2010}. The velocity dispersions of the PNe are consistent with the values derived by the stars kinematics and with those for the red GCs \citep{Napolitano2002,McNeil2010}.
The X-ray data have shown the presence of a hot gaseous halo associated to NGC~1399, which extends out to about 90 kpc \citep{Paolillo2002}. It appears very asymmetric, consisting of three components which dominate at different scales:  the central bright part, coincident with the peak of light, the 'galactic' halo that extends out to $\sim 7$ arcmin, which is offset toward the west with respect to the optical galaxy centroid, and,  on larger scales, the 'cluster' halo that is elongated in the opposite direction. The galactic halo is
very inhomogeneous, showing the presence of several cavities and arc-like features \citep[see Fig.3 and 6 in ][]{Paolillo2002}. 
Differently from the compact stellar systems and X-ray emission, the light distribution and colors in the faint outskirts of NGC~1399 are still unexplored. As part of the FDS, in this paper we present the analyzis of the deep photometry performed in the $g$ and $i$ bands inside two square degrees around NGC~1399 in order to unveil the stellar halo of this galaxy at the largest galactocentric distances and in unexplored range of surface brightnesses.

In the following, we adopt a distance for NGC~1399 of D=19.95 Mpc \citep{Tonry2001}, which yields an image scale of 96.7 parsecs/arcsec.

\section{Observations, Data Reduction and Analysis}\label{data}

The ongoing FDS observations are part
of the Guaranteed Time Observation surveys, {\it FOCUS}
(P.I. R. Peletier) and {\it VEGAS} \citep[P.I. M. Capaccioli,][]{Cap2015} being performed at the ESO VLT Survey
Telescope (VST).  VST is a 2.6-m wide field optical survey telescope,
located at Cerro Paranal in Chile \citep{Schipani2012}. VST is
equipped with the wide field camera OmegaCam, spanning a $1 \times
1$~degree$^2$ field of view, in the optical wavelength range from 0.3
to 1.0 micron \citep{Kui2011}. The mean pixel scale is
0.21~arcsec/pixel.

The observations presented in this work were collected during a 
visitor mode run from {  11 to 23 of November 2014 (runID: 094.B-0496[A]). Images are in the $u$, $g$, $r$ and $i$ bands. All images were acquired in dark time.} 
They cover five fields of 1~deg$^2$, one of them centerd
on the core of the Fornax cluster ($\alpha=03h38m29.024s$,
$\delta=-35d 27' 03.18''$).  Observations were obtained by using a {\it
step-dither} observing sequence, consisting of a cycle of
short exposures, one centerd on the core of the Fornax cluster and
others on 3 adjacent fields all around it ($\Delta = \pm 1$~degree)
on the SW and NW side. This approach, already adopted in other
photometric surveys \citep{Ferrarese2012}, allows a very accurate
estimate of the sky background around bright and extended galaxies.

{  For each field we have obtained 76 exposures (of 150~sec each) 
in the $u$ band, 54 in the $g$ and $r$ bands, and 35 in the $i$ band,
giving a total exposure time of 3.17 hrs in the $u$ band, 2.25 hrs in the $g$ and $r$ bands and  
and of 1.46 hrs in the $i$ bands}.  The
average seeing during the  observing run ranged between 0.6 and
1.1~arcsec.

The data reduction has been performed with the {\it VST-Tube} imaging
pipeline:  a detailed description of it, including calibrations and
error estimate, is given by \citet{Cap2015}.  The observing
strategy adopted for this data-set allows to derive an average sky
image, from exposures taken as close as possible in space and time, to the scientific
ones.  The average sky frame is derived for each observing night, then it is scaled and
 subtracted off from each science frame. 

\begin{figure*}[t]
%\epsscale{0.5}
%\plotone{fornax1_ugi_ex.pdf}
\includegraphics[scale=.8]{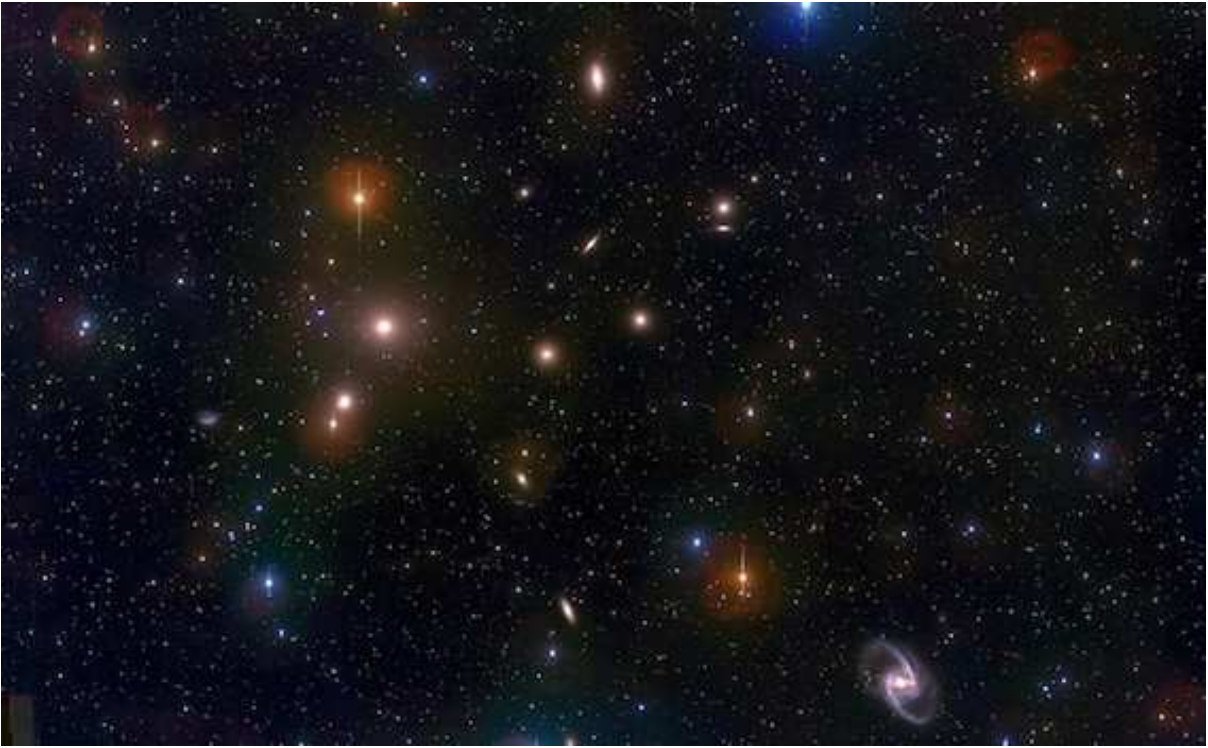}
\caption{\label{mosaic} Color composite mosaic (from the $u$, $g$ and
  $i$ bands) of the central $2 \times 1$~degrees$^2$ in the core of the Fornax cluster. The
  central cD galaxy NGC~1399 is the brightest object on the left side. North is up and East is on the
  left.} %Credits: L. Limatola and A. Grado.}
\end{figure*}

The color composite mosaic (from the $u$, $g$ and $i$ bands) of the
central $2 \times 1$~degrees$^2$ around the central cD galaxy NGC~1399
is shown in Fig.~\ref{mosaic}. In this work we concentrate our
analysis on the $g$ and $i$ band images of this region, in order to study the light distribution of the stellar halo and the diffuse ICL component.
In a forthcoming paper, where we will also present  the full dataset of the FDS observations, we will show  the photometry and structure of all the bright galaxies in the whole mosaic, including the 2-dimensional model of the light distribution (Iodice et al., in preparation).

\subsection{Surface Photometry}\label{method}

One of the goals of the FDS is to map the surface brightness and
color profiles out to the very faint outskirts of the cluster member galaxies (up to
about 8-10 effective radii $R_e$).  
{  To this aim, the crucial step is to account for all the effects that can contribute to the sky brightness with the highest accuracy. As mentioned in the previous section, data were acquired in completely dark nights.
By looking at the 100 micron image available at the NASA/IPAC Infrared Science Archive we have checked that the region of the Fornax cluster is almost devoid of cirrus, thus cirrus do not contaminate the images. Other possible sources that can contribute to the sky brightness are the extragalactic background light, the zodiacal light and the terrestrial airglow \citep{Bernstein2002}. The zodiacal light decreases in intensity with distance from the Sun, it is very faint at low ecliptic latitudes \citep{Bernstein2002}. Therefore, considering the ecliptic latitude of the Fornax cluster and the  low airmass value during the observations ($\le 1.5$), the contribution by the  zodiacal light and also of the terrestrial airglow are very low. 
Given that, the small contribution by the  smooth components listed above to the sky brightness is taken into account by the sky frame derived for each observing night and then subtracted off from each science frame (see Sec.~\ref{data}). Since each sky frame is taken on a field close in distance to the target, and taken close in time, the average value of all the smooth sources listed above is estimated in that way, thus only a possible differential component could remain in the sky-subtracted science frame. To estimate any residuals after the subtraction of the sky frame that can take into account also such a differential effects, we proceeded as follows.
On the whole sky-subtracted mosaic, in each band, after masking all the bright sources (galaxies and stars) and background
objects, we extracted the azimuthally-averaged intensity profile with the IRAF task ELLIPSE. 
The fit is performed in elliptical annuli, centerd on NGC~1399, with fixed Position Angle (P.A.) and ellipticity ($\epsilon$), given from the isophotes at $R \sim 1$~arcmin, where $P.A.=6$~degrees and $\epsilon = 0.02$. 
The major axis increases linearly with a step of 100 pixels out to the edges of the
frame, i.e. out to about 100~arcmin, excluding the border that is characterised by a larger noise.
This method allows us {\it i)} to measure any residual fluctuations\footnote{  The Òresidual fluctuationsÓ in the sky-subtracted images are the deviations from the the background in the science frame with respect to the average sky frame obtained by the  empty fields close to the target. Therefore, by estimating them, we obtain an estimate on the accuracy of the sky-subtraction step.} in the sky-subtracted images and then take them into account in the error estimate of the surface brightness, and {\it ii)} to estimate the distance from the galaxy center where the galaxy's light blends into the background at zero counts per
pixel, on average. This method was firstly suggested by \citet{PhT06} to study the light profiles of spiral galaxies and also adopted by \citet{Iod2014} on the deep VISTA wide-field images of NGC~253.}

\begin{figure*}[t]
%\epsscale{1.1}
%\plottwo{sky_conf1.ps}{conf_profg.ps}
\includegraphics[scale=.4]{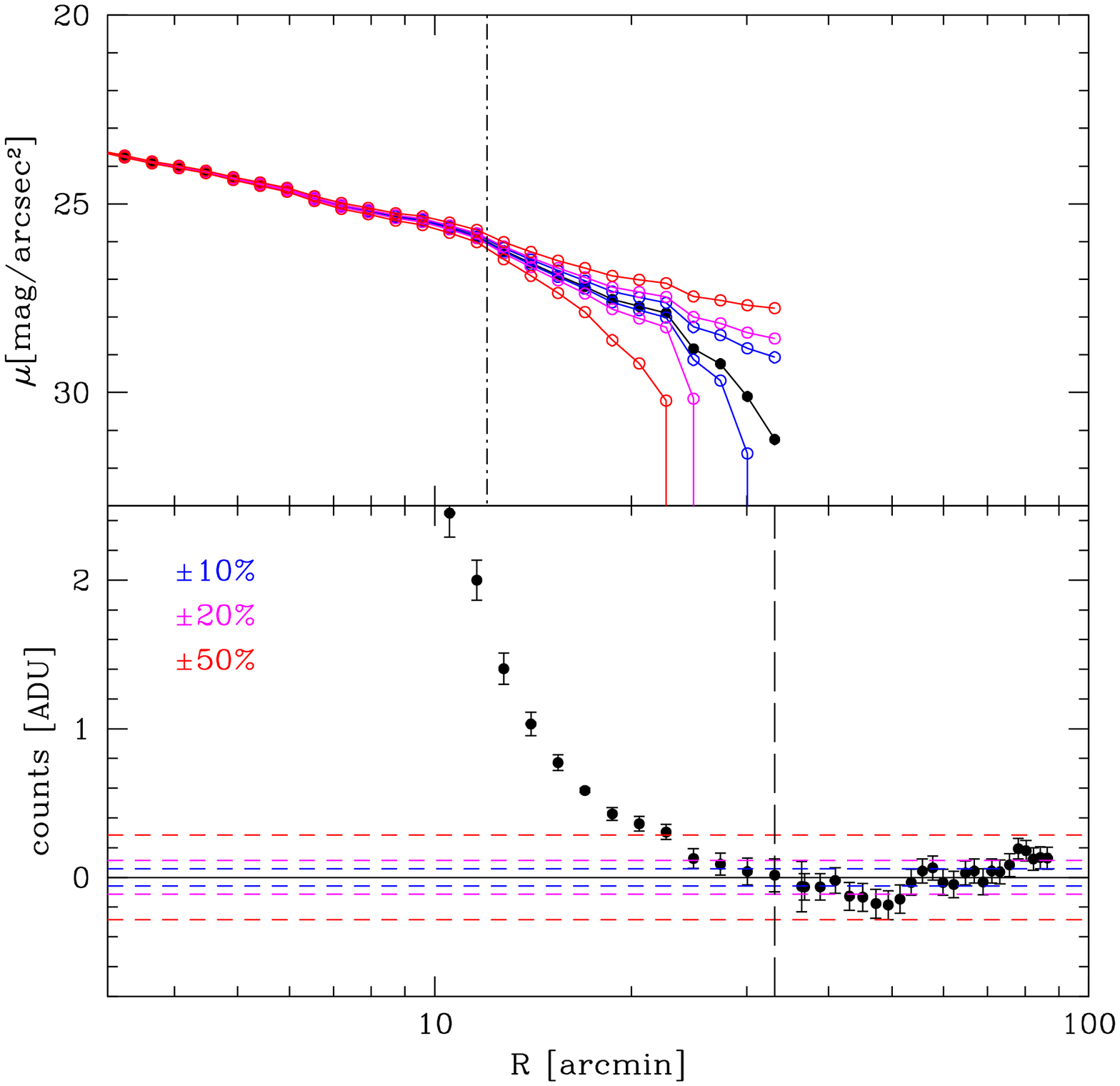}
\includegraphics[scale=.4]{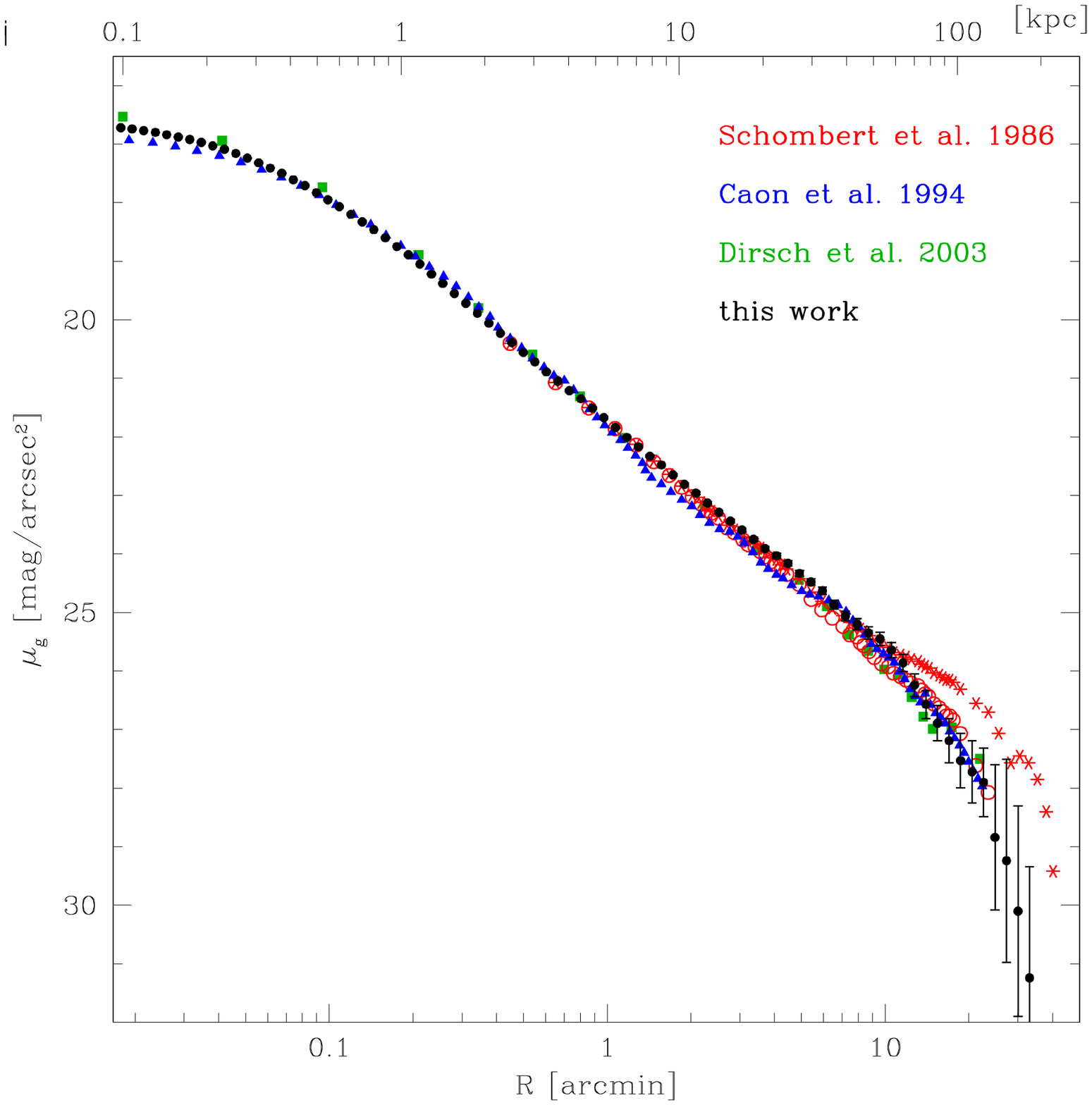}
\caption{\label{sky} {\it Left panel} - Surface brightness profile
  (top) and intensity profile (bottom) in the g band, in the outer
  regions of NGC~1399. The black points correspond to the values
  corrected for the background level $\pm 0.05$~counts (continuous
  black line), estimated at $R \ge R_{lim}$ (long dashed line), see bottom panel. The
  open blue, magenta and red points correspond to the surface
  brightness values for a sky levels of $10\%$, $20\%$ or $50\%$ lower
  or higher than our estimate of the background level, respectively. The dot-dashed vertical
  line indicates the break radius at $R=12$~arcmin ($\sim 70$~kpc),
  see text for details. In the bottom panel, the short dashed lines
  indicates the corresponding counts levels of the residual sky for
  the above values. {\it Right panel} - Total surface brightness
  profile in the g band (black circles) derived from VST mosaic,
  compared with literature data (top panel).  The blue triangles are
  magnitudes derived from \citet{Caon1994},  {  green} squares are from \citet{Dirsch2003}, the
  red asterisks are the original data from \citet{Schombert86} and the open red circles are the same magnitudes corrected for a 30\%  higher value of the sky level.}
\end{figure*}

From the intensity profiles, we estimated the outermost radius, from
the center of the galaxy, where counts are consistent with those from
the average background level. Such a value is the residual by the subtraction of the sky frame,  thus it is very close to zero.
The radius where the galaxy's light blends into this level sets the surface brightness limit of the VST light profiles  and gives an
estimate on the accuracy of the sky subtraction. 
In the bottom panel of Fig.~\ref{sky} we show the  intensity profile in the $g$ band, centerd on NGC~1399 and
extending out to the edge of the mosaic, as a function of the
semi-major axis. We derived that for $R \ge 33$~arcmin ($\sim 192$~kpc) we
are measuring the residual fluctuations of the sky. The error on the average value is 0.11 counts. The fluctuations of the background level are less than 0.2 counts. 
The same approach is adopted in the $i$ band and it gives a
limiting radius of $R \sim 27$~arcmin, an error on the average value of 0.2 and fluctuations of the background level that are less than 0.4 counts.  

\begin{figure*}[t]
%\centering
%\epsscale{1.1}
\includegraphics[scale=.4]{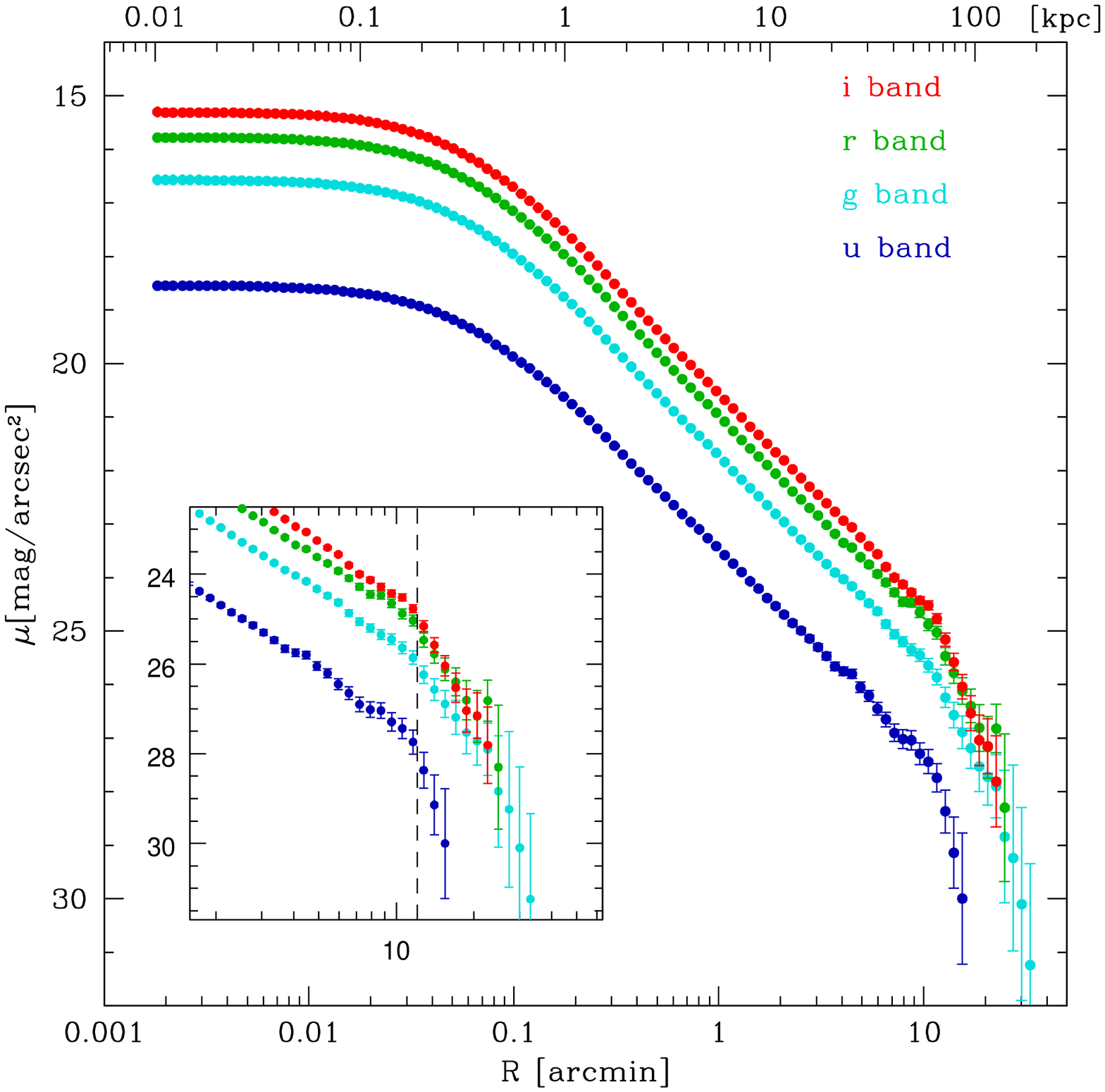}
\includegraphics[scale=.4]{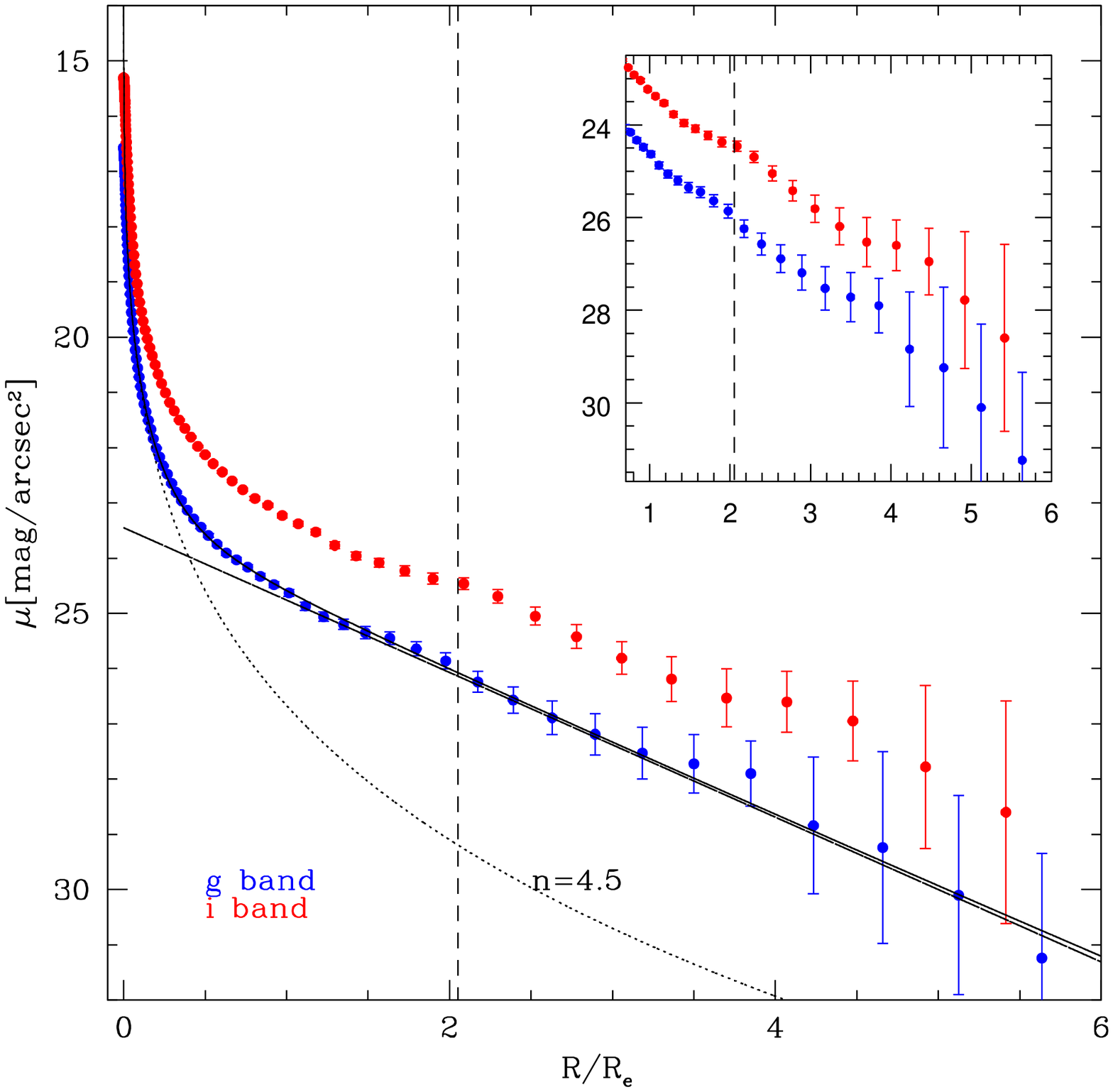}
\caption{\label{profiles}   {\it Left panel -} Azimuthally averaged surface brightness
  profiles in the $u$ (blue), $g$ (light blue), $r$ (green) and $i$ (red) bands from the VST mosaic in logarithmic scale.  {\it Right panel -} $g$ and $i$ bands profiles plotted in linear scale. The effective radii are those derived by fitting the growth curves of the $g$ and $i$ total magnitudes. The continuous line is  the resulting fit with an exponential (long-dashed line) and a Sersic (dotted line) law (see Sec.~\ref{fit}).
  In both panels, the small box is the enlarged  portion of the outer regions where the light profiles become steeper  for $R \ge 12$~arcmin ($\sim 70$~kpc). This radius is indicated
  with the vertical short-dashed line.}
\end{figure*}

\begin{figure}[t]
%\epsscale{1}
%\plotone{prof_axis.ps}
\includegraphics[scale=0.35]{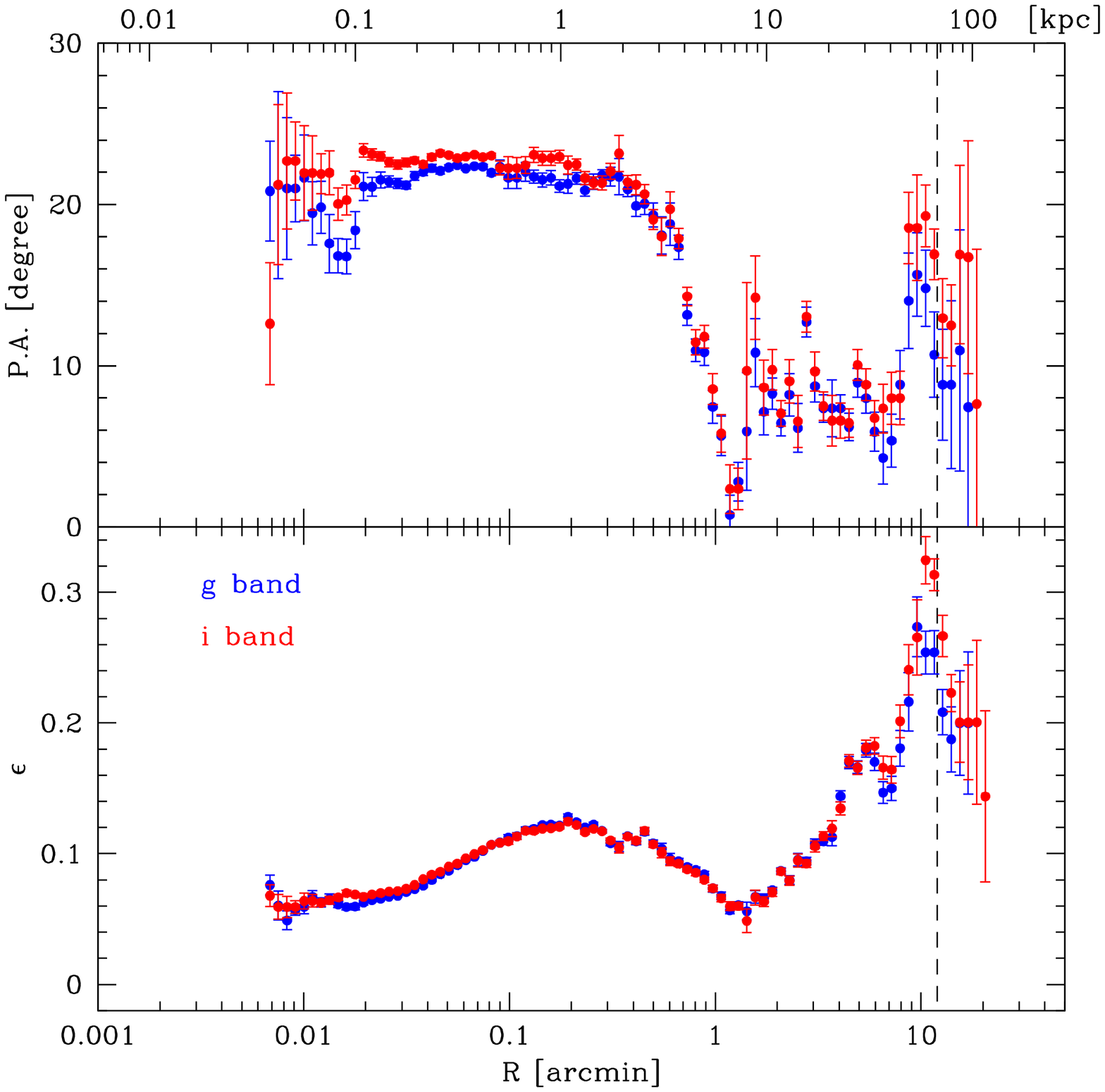}
\caption{\label{epsPA} Ellipticity (bottom panel) and P.A. (top panel) profiles for NGC~1399 derived by fitting the isophotes in the $g$ (blue points) and $i$ (red points) band images. The vertical dashed line indicates the break radius at $R = 12$~arcmin.}
\end{figure}

The isophotal analysis of NGC~1399  is performed by using the ELLIPSE task out to the limiting radii
estimated above, where  the semi-major axis length is sampled with a variable bin. 
We derived  the azimuthally averaged surface brightness profiles (see Fig.~\ref{sky} and Fig.~\ref{profiles}), as well as the ellipticity and P.A. profiles (see Fig.~\ref{epsPA}), as function of the semi-major axis.
The surface brightness profiles extend down to  $\mu_g = 31 \pm 2$~mag arcsec$^{-2}$ and
$\mu_i = 28.6 \pm 2$~mag arcsec$^{-2}$, in the $g$ and $i$ bands
respectively.  According to \citet{Cap2015} and \citet{Seigar2007}, the error estimates\footnote{The uncertainty in the surface brightness is calculated with the following formula: $err= \sqrt{(2.5/(adu \times \ln(10)))^2 \times ((err_{adu}+err_{sky})^2) +  err_{zp}^2}$, where
$err_{adu}=\sqrt{adu/N-1}$, with N is the  number of pixels used
in the fit, $err_{sky}$ is the rms on the sky background and $err_{zp}$
is the error on the photometric calibration \citep{Cap2015,Seigar2007}. }  
on the surface brightness magnitudes take the uncertainties on the photometric calibration
($\sim 0.003 - 0.006$~mag) and sky subtraction ($\sim 0.11 - 0.2$~counts) into account.
{  We noticed that, for a large area of the mosaic ($\sim 50$ to $\sim80$~arcmin from the galaxy center) the fluctuations on the background level are not larger the 10\% (see bottom panel of Fig.~\ref{sky}). While, on a smaller range of radii, for $30 \leq R \leq 40$~arcmin and for  $80 \leq R \leq 90$~arcmin from the galaxy center, the deviations from the average zero-level of the background  affect by about the 20\% the surface brightness  for  $\mu_g \ge 27$~mag~arcsec$^{-2}$ and $\mu_i \ge 25$~mag~arcsec$^{-2}$. Therefore, to adopt a conservative approach the error estimate on the sky subtraction takes into account the largest values of about 20\% in the residuals, even if it is worth  stressing that this is an upper limit and that this value is smaller, on average.}
 
Before discussing the light profiles and the 2-dimensional light and
color distribution (see next section), we present here some tests on
the reliability of our results.

We noticed that the light profiles show a change in the slope at
$R=12$~arcmin ($\sim 70$~kpc) and at $26 \le \mu_g \le
26.5$~mag~arcsec$^{-2}$ (see Fig.~\ref{profiles}). In a  logarithmic scale this is quite evident. At this {\it break
  radius} the light profiles become steeper.  The relatively high surface brightness level 
  at which the change of slope is observed suggests that it should not be due to wrong 
  estimate of the sky level. Nevertheless, we have checked this as follows.
In the left panel of Fig.~\ref{sky} we show how much the light
profile changes by accounting for a sky level $10\%$, $20\%$ and
$50\%$ higher or lower the average background level. Even if the average sky level was wrong by 
$20\%$ the change in the slope would be still present and only the outer
points (at $R \ge 25$~arcmin) would be affected. 
%Both a $10\%$ and a $20\%$ lower value of the sky level is consistent with the error estimate on the intensity (see bottom panel of Fig.~\ref{sky}). 
To obtain a shallower profile from $R \ge 20$~arcmin, of about more than
2 mag~arcsec$^{-2}$, the sky level would have to be  $50\%$ lower. Anyway,
also in this case, for $12 \le R \le 20$~arcmin a slope variation in the
light profile is still evident.  By looking at the bottom panel of the
Fig.~\ref{sky}, this large difference can be excluded since the
observed background level out the edge of the mosaic is clearly
higher. 

A further check comes by comparing the light profiles obtained from
the VST data with literature data for NGC~1399. All the light profiles from the literature were shifted to the $g$-band profile, by using the transformations from \citet{Fukugita1996}.
The most extended light profiles published in the optical bands for NGC~1399 are from \citet{Schombert86}, \citet{Caon1994} and \citet{Dirsch2003} and they are compared with the VST profile in the $g$ band in the right panel of Fig.~\ref{sky}. On average, the agreement is
quite good.
The light profile by \citet{Caon1994} extends out to 14~arcmin. It was obtained along the major axis of NGC~1399 in the B band and this causes some differences, between the two profiles, that are evident for $1 \le R \le 6$~arcmin. A change in the slope is observed at the same radius ($R=12$~arcmin) as the one derived from the VST $g$-band profile (see right panel of Fig.~\ref{sky}). 
The light profile published by \citet{Dirsch2003} (in the R band) is not an azimuthally average but it is derived by measuring the surface brightness around all targeted GCs sources within an annulus between 8 and 11 arcsec.  Compared to the VST profile, the agreement is very good. Even if it is less extended and does not reach the  same faint surface brightness levels as our data, a drop is still observed for $R \ge 10$~arcmin.
The most extended light profile ($R\sim40$~arcmin) available in literature is that from \citet{Schombert86} in the V band. It is consistent with the VST profile out to $R\sim 10$~arcmin. As previously argued by \citet{Dirsch2003}, the discrepancy of this profile from all others  might be due to an overestimated background and/or additional light from the bright objects close to NGC~1399, that were not completely  masked. Indeed, if we correct the background level for the profiles published by  \citet{Schombert86} of about the $30\%$, it becomes fully consistent with the other results (see open red circles in the right panel of Fig.~\ref{sky}).

\section{Results: Light and color distribution out to 192 kpc}\label{result}

\begin{figure*}
%\centering
\includegraphics[scale=0.55]{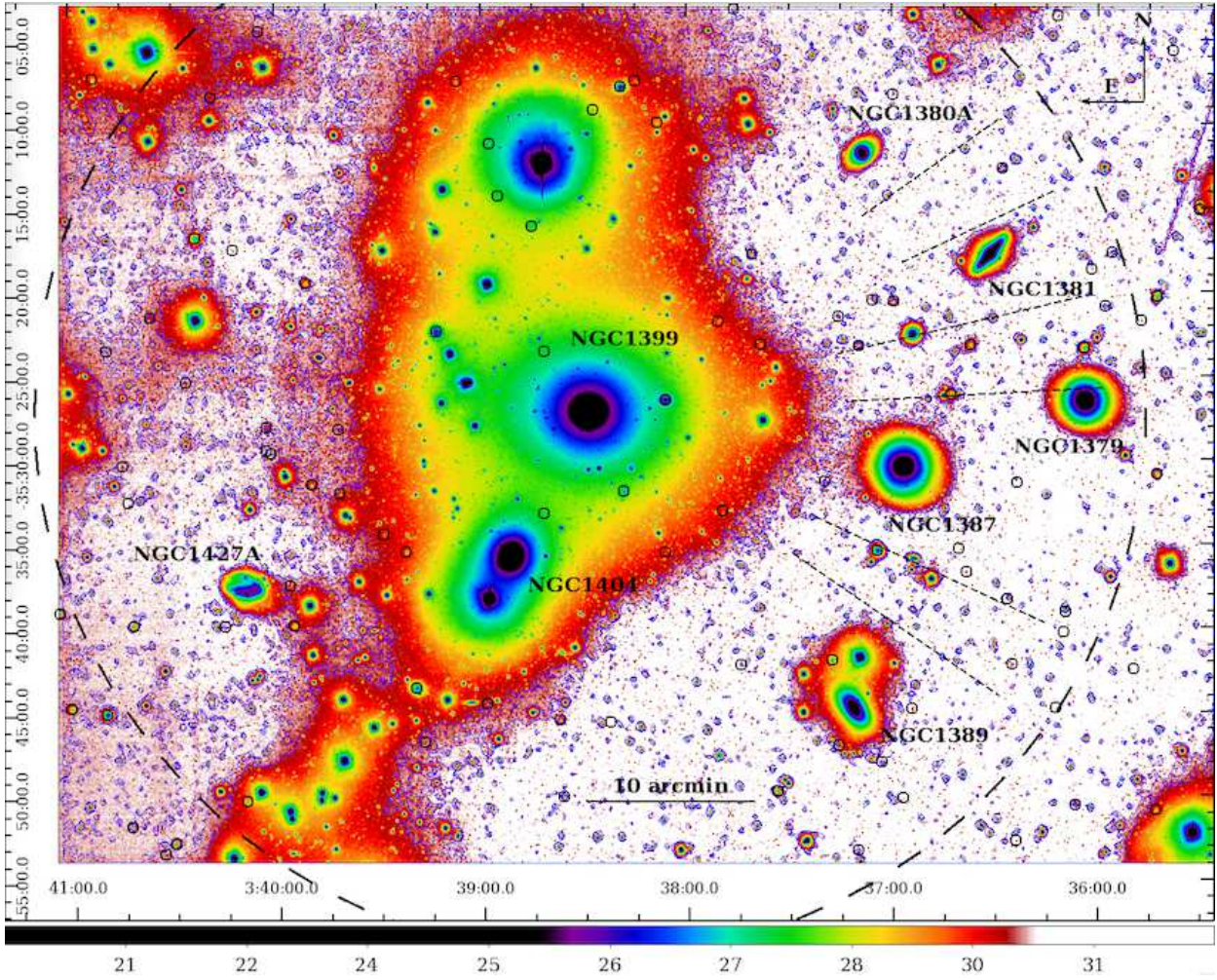}
\caption{\label{1399halo}  VST image  centered on NGC~1399 in the $g$ band, plotted in surface brightness
  levels. The image size is $68.95 \times 51.10$~arcmin. The dashed black ellipse is the isophote at the limiting radius  $R_{lim}= 33$~arcmin ($\sim 192$~kpc). The surface brightness level contours between $30 \le \mu_g \le 31.2$ mag~arcsec$^{-2}$ are shown in blue. The dashed black lines direction along which we extracted the radial profiles shown in Fig.~\ref{axis} and we observe a light excess. {  The black circles  identify the whole sample of dwarf galaxies  in the Fornax cluster observed in the DECAM images published by \citet{Munoz2015}.}}
\end{figure*}

In Fig.~\ref{1399halo} we show the {  the sky-subtracted} VST image in the
$g$ band centerd on NGC~1399, plotted in surface brightness levels,
where the outermost isophote at $R=33$~arcmin ($\sim 192$~kpc) is overlaid. {  In order to highlight the low surface brightness structures we performed a gaussian smoothing with a radius of 20 pixels.}

\begin{figure*}[t]
%\cantering
\includegraphics[scale=0.8]{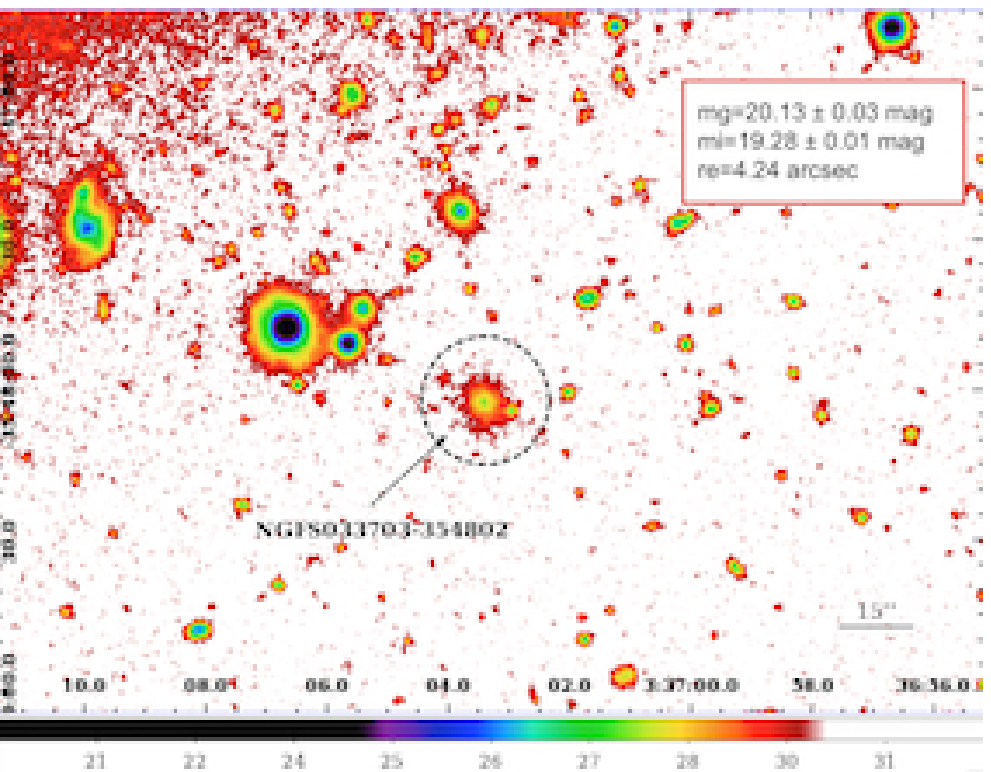}
\includegraphics[scale=0.8]{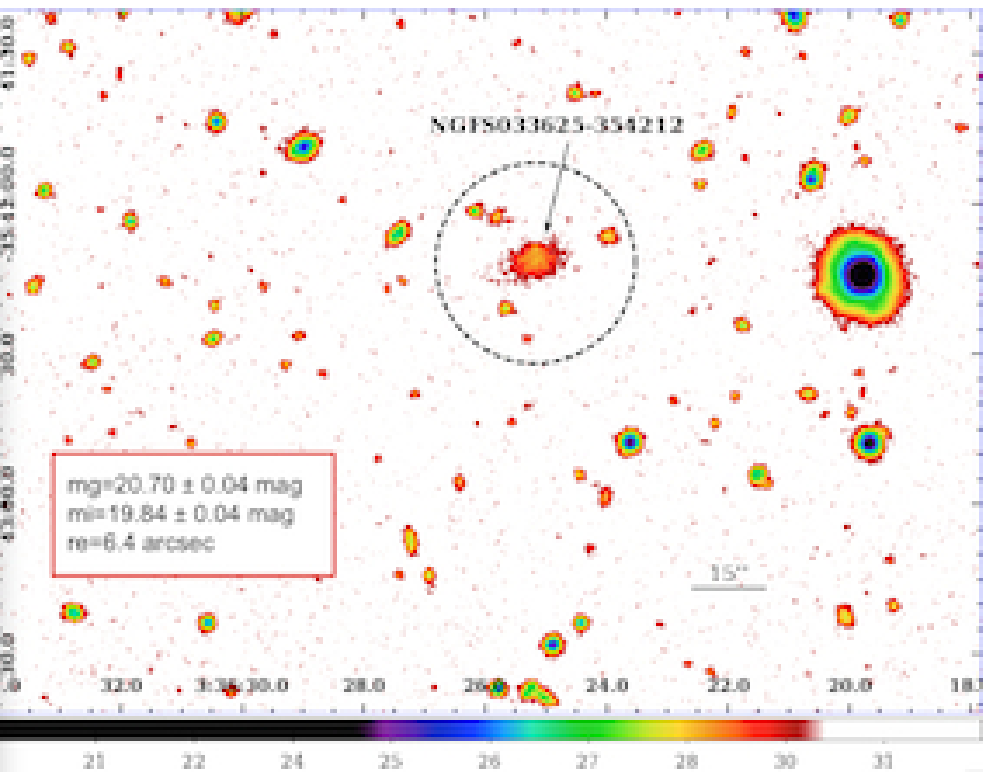}
\caption{\label{dwarf}   Enlarged portion of the mosaic where two dwarf galaxies among the faintest objects detected by \cite{Munoz2015} are located. We selected two examples, one nucleated (left panel) and one non-nucleated (right panel) dwarfs. The integrated magnitudes in the $g$ and $i$ bands were derived in a circular aperture with a radius of $2 r_e$, where the effective radius $r_e$ is the value given by \cite{Munoz2015}.}
\end{figure*}

At these large distances we are mapping the faint stellar
outskirts  of NGC~1399, which encloses most of the bright galaxies in the core of the Fornax cluster.  
{  On this image we have identified the whole sample of the dwarf population in the Fornax cluster observed in the DECAM images, published by \citet{Munoz2015}.  As an extra-check of the detection limit  of the VST data, we provide in Fig.~\ref{dwarf} the enlarged portion of the mosaic in the $g$ band where two dwarf galaxies among the faintest and new objects discovered by \cite{Munoz2015} are located. We selected two examples, one nucleated and one  non-nucleated dwarf, which appear clearly detectable also in the VST. }

\begin{figure*}
%\centering
\includegraphics[scale=0.5]{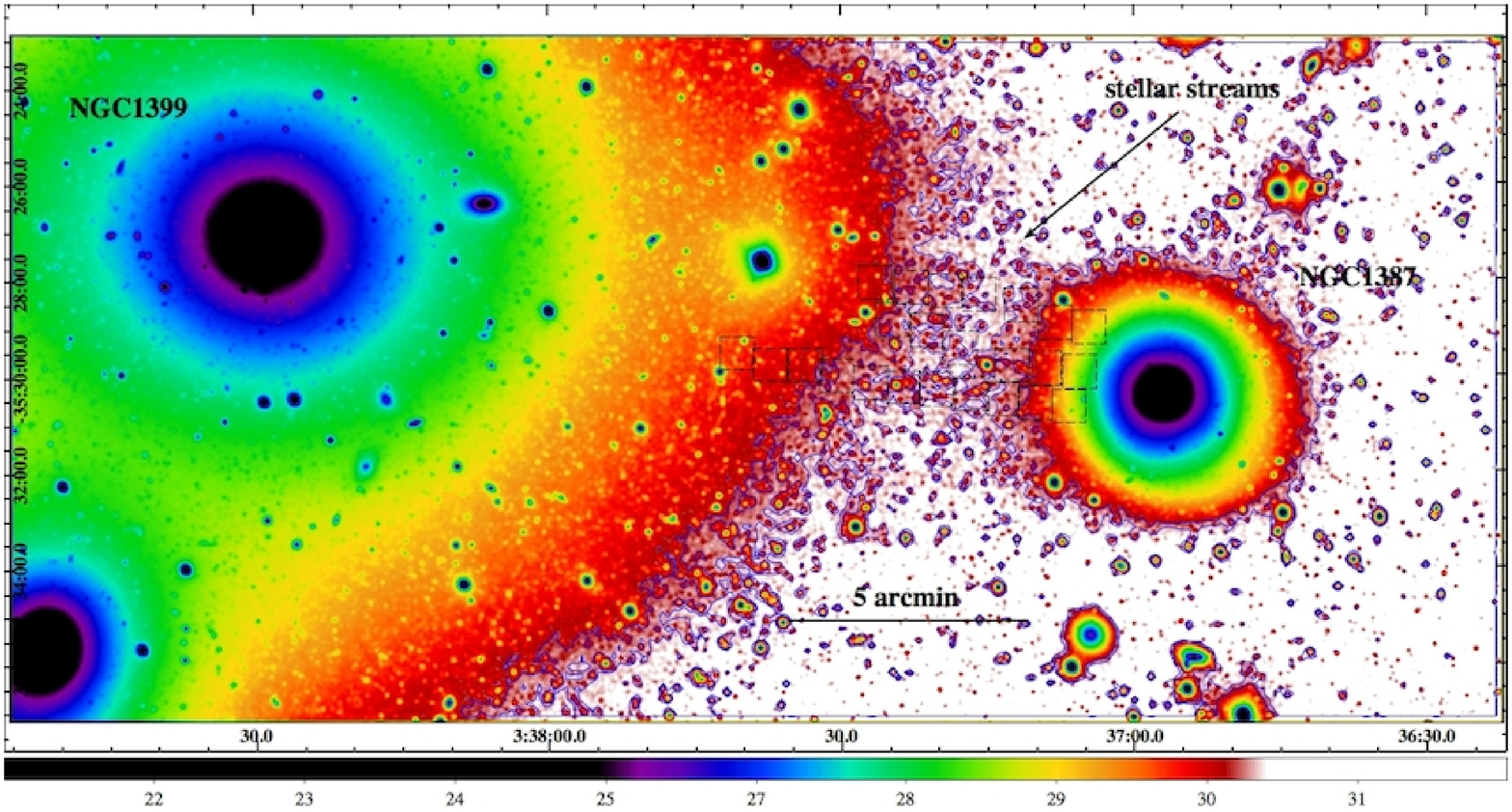}
\includegraphics[scale=0.5]{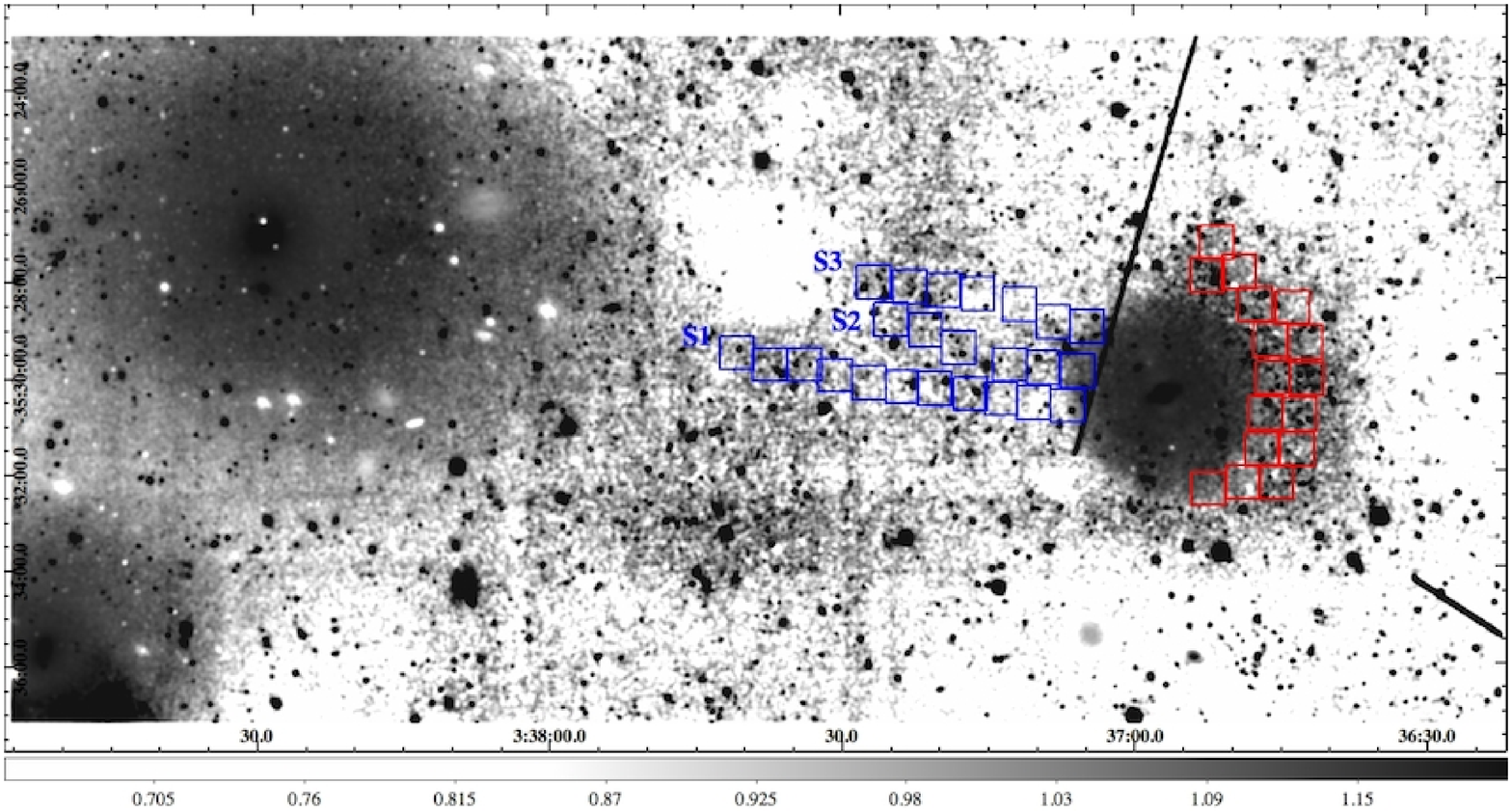}
\caption{\label{streams}   {\it Top panel -} Enlarged image of the core of the Fornax cluster in the VST $g$ band shown in Fig.~\ref{1399halo}, on the West regions of NGC~1399.  The arrow indicates the regions of the streams between NGC~1399 and NGC~1387. Dashed boxes correspond to  the regions where an average $g-i$ color is derived. {\it Bottom panel -} $g-i$ color map in the same region shown in the top panel. Darker colors correspond to redder $g-i$ values. Boxes correspond to the regions where an average $g-i$ color is derived for the streams (blue) and in the halo of NGC~1387 (red). The image size is $31.1 \times 14.3$~arcmin.}
\end{figure*}

{  \subsection{The low surface brightness substructures in the outskirts  of NGC~1399 } 
In the region between NGC~1399 and NGC~1387 we detect a "bridge-like"  stellar stream. This feature is much more evident in the enlarged image shown in Fig.~\ref{streams} (top panel): it appears to be made by several filamentary structures  that  connect the two galaxies. They are about 5~arcmin long ($\sim 29$~kpc) and  cover an area of about 4.5~arcmin ($\sim 27$~kpc). They have an average surface brightness of $\mu_g \sim 30$~mag~arcsec$^{-2}$ ($\mu_B\simeq 30.6$~mag arcsec$^{-2}$). 
%Besides, we noticed that, at these faint levels of surface brightnesses the isophotes of NGC~1399  appear "rugged" and more elongated toward the west side. There are many of these kind of  stellar filaments but less extended (0.5 - 2~arcmin) than those observed between NGC~1399 and  NGC~1387. 
Also the isophotes of NGC~1387 are not symmetric with respect to the galaxy center, appearing more elongated toward the SW.

The $g-i$ color map in the area of the stellar streams, shown in the bottom panel of Fig.~\ref{streams}, shows that, on average, the stellar bridge is  bluer than the two close galaxies. The asymmetry in the SW of NGC~1387 is a very extended arc-like region with redder colors.
We derived the average  $g-i$ colors in several regions covering  the three main filaments in the stellar bridge (see boxes in Fig.~\ref{streams}) and in the red arc on the west side of NGC~1387  (see red boxes in the bottom panel of Fig.~\ref{streams}). Results are plotted in Fig.~\ref{stream_col}.
We found that in the middle region of the stellar bridge the $g-i \sim 0.9$~mag. This value is about $0.1 - 0.2$~mag bluer than the average colors in the regions closer to NGC~1399 and NGC~1387, which are about 1.08 and 1.14 mag, respectively (see Fig.~\ref{stream_col}).
Given the very low level of the surface brightness, the stellar bridge and the features identified in the color map were unknown in the previous works. 

The existence of a stellar stream towards NGC~1387 was already proposed by \citet{Bassino2006}  
based on their study of the projected distribution of the blue GCs (see Fig.9 of that paper) and recently confirmed by \citet{Dabrusco2016} using our FDS data (see Fig.~3 and Fig.~4 of that paper).   
The system of stellar streams detected in this work match very well in space and size the bridge detected in the blue GCs spatial distribution.

Differently from the GCs, the spatial distribution of the dwarf galaxies do not show any evident correlation with the stellar stream between NGC~1399 and NGC~1387 (see Fig.~\ref{1399halo}). 
This cannot be considered as a definitive conclusion since we cannot exclude the presence of undetected faint and diffuse dwarfs inside the stellar streams (this aspect will be further investigated in a dedicated work on the FDS data).

The stellar bridge could be the sign of an ongoing interaction between the two galaxies. 
This aspect is discussed in detail in Sec.~\ref{discussion}.}

%\begin{figure*}
%\centering
%\includegraphics[scale=.4]{streams.eps}
%\caption{\label{streams} Enlarged image of the core of the Fornax cluster in the VST $g$ band shown in Fig.~\ref{1399halo}, on the West regions of NGC~1399.  The arrow indicates the regions of the streams between NGC~1399 and NGC~1387.}
%\end{figure*}

\subsection{Isophote analysis}\label{ellipse}
{  The azimuthally averaged light profiles of NGC~1399 in the $u$, $g$, $r$ and $i$ bands are shown in the left panel of Fig.~\ref{profiles}. The light profile in the $g$ band is the most extended,  out to 33~arcmin ($\sim 192$~kpc) from the galaxy center, and the deepest,  down to $\mu_g = 31 \pm 2$~mag arcsec$^{-2}$ ($\mu_B\simeq 31.8$~mag arcsec$^{-2}$).
In the $i$ band, we reach the faint level of $\mu_i = 29 \pm 2$~mag~arcsec$^{-2}$ at $R\sim 27$~arcmin (see Fig.~\ref{profiles}).  
As pointed out in the previous section, the $g$-band light profile shows a  slope variation at $R_{break}=12$~arcmin $\sim 2R_e$ ($\sim 70$~kpc) and at $26 \le \mu_g \le 26.5$~mag~arcsec$^{-2}$ ($26.6 \le \mu_B \le 27.1$~mag arcsec$^{-2}$), which is quite evident when plotted logarithmically. This is a common feature in the light profiles of all  other bands, where for $R \ge R_{break}$ they are steeper.}

By integrating the growth curve of the azimuthally averaged profiles,  the total magnitudes inside 33~arcmin in the $g$ band and 27~arcmin in the $i$ band are $m_g=8.57\pm 0.14$~mag and $m_i=7.53 \pm 0.10$~mag {  ($m_B \simeq 9.17$~mag)}, respectively. Values were corrected for the Galactic extinction by using the absorption coefficient $A_{g}=0.035$ and $A_{i}=0.018$ derived according to \citet{Schlegel98}. 
{  The absolute magnitude in the $g$ band is $M_g=-22.93$~mag, therefore the total luminosity is $L_{g} \simeq 1.66 \times 10^{11} L_{\odot}$.} 
The effective radii are $R_e=5.87 \pm 0.10$~arcmin, in the $g$ band, and $R_e=5.05 \pm 0.12$~arcmin and $i$ band.  
The above values for the effective radii are larger than that given by \citet{Caon1994}, which is $R_e = 2.17$~arcmin: this is the effect of a more extended light profile in the VST $g$-band images (see right panel of Fig.~\ref{sky}). We derived the effective radius integrating the grow curve out the last data point in the \citet{Caon1994} profile, i.e. $R=14$~arcmin, and we have obtained $R_e = 2.28$~arcmin, which is consistent with the already published value.  
According to the $R_e$ estimates given above, the surface brightness profiles in $g$ and $i$ bands extend out to about $6 R_e$ (see right panel of Fig.~\ref{profiles}).  

The Position Angle (P.A.) and ellipticity ($\epsilon = 1 - b/a$, where $b/a$ is the axial ratio of the ellipses) of the isophotes change towards larger radii (see Fig.~\ref{epsPA}). Inside 0.5~arcmin, isophotes are almost round ($0.06 \le \epsilon \le 0.12 $) and the P.A. is constant at about 22 degrees. For $0.5 \le R \le 1$~arcmin a large twisting is observed: the P.A. decreases by about 20 degrees. At larger radii, for $R \ge 1$~arcmin, the ellipticity increases and reaches a maximum of $\epsilon \sim 0.25$ in correspondence of the break radius, while the P.A. remains constant at $\sim 8$~degrees and it also shows a peak ($P.A. \sim18$~degrees) at $R=R_{break}$. 
Thus, for $R \ge 1$~arcmin, the light distribution tends to be more flattened and elongated towards the SW direction than in the inner regions.

\subsection{ Light profiles at several P.A.s} 
The steeper behaviour of the outer profile is confirmed by the light profiles extracted along several position angles: at P.A.=98~degrees, which correspond to the major axis of the outer component (for $R\ge R_{break}$, see Fig.~\ref{epsPA}), and other 3 intermediate directions at P.A.=30~degrees, P.A.=40~degrees and P.A.=120~degrees. The surface brightness is computed in a sector of ten degrees wide and corrected for the flattening. Results are shown in Fig.~\ref{axis}.
Profiles on the NE side at P.A.=30~degrees and on the SE side at P.A.=30~degrees are not included in the plot since most of the regions out to 5~arcmin fall in the masked areas. 

We found that the light distribution of the fainter outskirts at $R\ge R_{break}$ has on average the same light profile at different position angles. In particular, even taking into account the larger scatter than the azimuthally averaged profile, at all P.A.s the light profiles show a change in the slope for $R\ge R_{break}$. 
On the West side, for $15 \le R \le 30 $~arcmin an excess of
light is measured at P.A.$=98$~degrees, where $27 \le \mu_g \le
28$~mag~arcsec$^{-2}$ {  ($27.6 \le \mu_B \le 28.6$~mag arcsec$^{-2}$)}, which is two magnitudes brighter than the light at the correspondent
radii on the East side. This region is  the intracluster area 
among NGC~1387, NGC~1379 and NGC~1381, where also \citet{Dabrusco2016}  find an excess of intracluster GCs. On the same side, at
P.A.$=120$~degrees (NW) a peak in the surface brightness is observed
at $R\sim20$~arcmin, with $\mu_g \sim 28$~mag~arcsec$^{-2}$ {  ($\mu_B\simeq 28.6$~mag arcsec$^{-2}$)}: on the
image, this is the region between NGC~1381 and NGC~1380B. Similar 
features are observed on the SW side, at P.A.$=40$~degrees, with $27
\le \mu_g \le 28$~mag~arcsec$^{-2}$ {  ($27.6 \le \mu_B \le 28.6$~mag arcsec$^{-2}$)} for $15 \le R \le 20$~arcmin, almost
one magnitude brighter than the NW profile at the same radii: at this
distance from NGC~1399 we are mapping the light between NGC~1389 and
NGC~1387.  
As discussed in Sec.~\ref{method}, at these faint surface brightness, small fluctuations of the
residual sky level could account for such kind of effect, which might become even more significant 
on a smaller area than a large elliptical annulus.  
Even if no definitive conclusion can be drawn at this
point, since we are mapping regions where the surface brightness
values are comparable with the sky fluctuations, it is important to
note that such "excess" of light is measured on the West side of the cluster core where most of the galaxies are located, while, on the East side, inside the same radius,  only the
dwarf irregular galaxy NGC~1427A is present (see Fig.~\ref{mosaic} and
Fig.~\ref{1399halo}).

In a forthcoming paper we show the analysis of the light distribution for all the bright galaxies in the core of the Fornax cluster, therefore, after modelling it and subtract off from the whole image, we could address wether those light excesses are still present in this regions or, alternatively, they are just the overlapping stellar halos of these galaxies.

\begin{figure}[t]
%\centering
\includegraphics[scale=0.4]{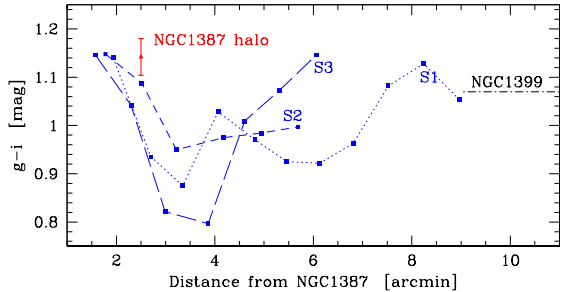}
\caption{\label{stream_col}   Average $g-i$ colors derived in the regions covering the streams (blue points) between NGC~1399 and NGC~1387 (see Fig.~\ref{streams} plotted as function of the distance from the center of NGC~1387. The red point indicates the average value of the $g-i$ color in the halo of NGC~1387. For comparison, the average value of $g-i$ derived for NGC~1399 at 10 arcmin from its center is also shown.}
\end{figure}

\begin{figure}[t]
%\epsscale{1}
%\plotone{prof_axis.ps}
\includegraphics[scale=0.4]{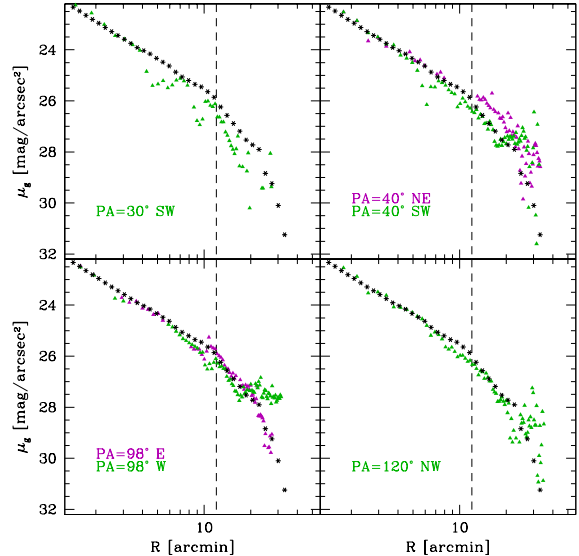}
\caption{\label{axis} Surface brightness profiles in the $g$ extracted along several P.A.s (green and magenta triangles) compared with the azimuthally averaged profile (black asterisks). The vertical dashed line indicates the break radius at $R = 12$~arcmin.}
\end{figure}

\subsection{ Color profiles and color map }
\begin{figure}[t]
%\epsscale{1.1}
\includegraphics[scale=0.4]{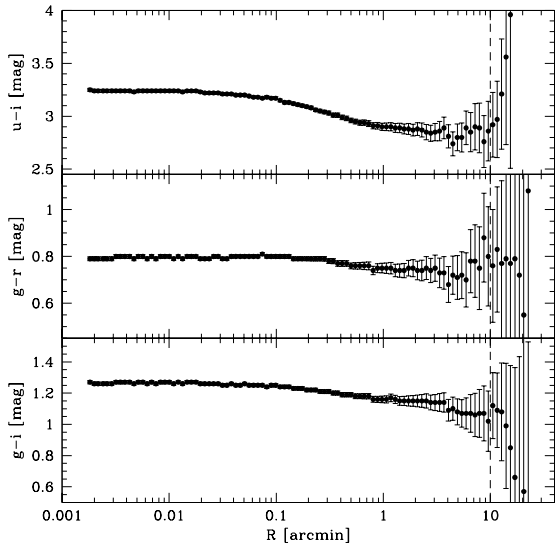}
\caption{\label{color}    Azimuthally averaged color  profile as function of the radius. The dashed line indicates the break radius  $R = 12$~arcmin ($\sim 70$~kpc). }
\end{figure}

\begin{figure}[t]
%\epsscale{1.1}
\includegraphics[scale=1.5]{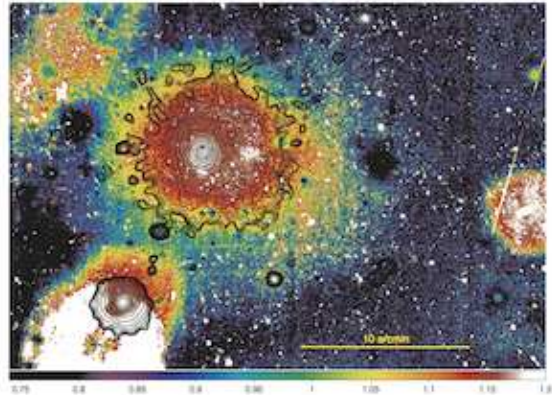}
\caption{\label{color_map}    $g-i$ color map centered on NGC~1399.  The image size is $31.24 \times 21.22$~arcmin.  The continuous black contours are the X-ray levels from \citet{Paolillo2002}. }
\end{figure}

{  The $u-i$, $g-r$ and $g-i$  color profiles are shown in Fig.~\ref{color}. 
The center of the galaxy is red, with $g-i \sim 1.3$~mag and $g-r \sim 0.8$~mag. A strong gradient towards bluer values is observed in the $u-i$ color profile, where  $3.2 \le u-i \le 2.7$~mag for $1 \le R \le 10$~arcmin.
In the same range of radii, the difference in the $g-r$ and the $g-i$ values is only $\sim0.1$~mag.
Close to the break radius, all the color profiles show a bump of about 0.1-0.2 magnitude redder than the central values.

For $R > R_{break}$,  $g-r$ has almost the same value as the  inner radii, i.e. $g-r \sim 0.8$.
The $g-i$ colors seem to be bluer, since $1.3 \le g-i \le 1.1$~mag, while the $u-i$ are redder up to 1 magnitude. 
This effect can  be explained as a gradient in metallicity \citep[see e.g.][]{Peletier1990}. 
%Such a difference could be due to the lower efficiency of the VST filter in the $u$ band \citep{Kui2011}, therefore at these faint levels of surface brightness we are missing flux in this band. 
%Taking into account that, in this region, the errors affecting the colors are quite large, color profiles are all consistent.

In Fig.~\ref{color_map}, we show the $g-i$ color map centered on NGC~1399.
In the inner regions, at $R \sim 3$~arcmin from the center, there is an extended red tail ($1.1 \le g-i \le 1.3$~mag) on the West side of the galaxy. This could be a sign of a recent minor merging event. 
At the same distance from the center ($R\sim 3$~arcmin) and on the same side (west), \citet{Coccato2013} found asymmetries in the kinematics traced by GCs:  a recent accretion of a dwarf galaxy was given as a possible explanation of this feature.
For $R \le R_{break}$, the color map appears asymmetric with respect to the center, it is more extended and redder on the SW side, where $g-i \sim 1.0 - 1.2$~mag. 
By comparing the X-ray density contours to the $g-i$ color map, we found a surprising correspondence
between the morphology and extension of the 'galactic' X-ray halo \citep[see][]{Paolillo2002} on
the west side of the galaxy and the color asymmetries observed in the stellar light. In particular, X-ray contours match with the shape of the red tail observed in the inner regions of the galaxy on the west side.}

\subsection{Profile fitting: the exponential halo around NGC~1399}\label{fit}

As pointed out in Sec.~\ref{intro}, several recent works have shown that the light profiles of  BCGs are not  uniquely and well fitted by a single Sersic profile, where the intensity $I(R)\propto R^{1/n}$, but in several cases a second component is needed to fit the outer stellar halo \citep{Seigar2007,Donzelli2011,Arnaboldi2012}. 
For NGC~1399, \citet{Schombert86} tried to fit the light profile with a single component, using the de Vaucouleur's law: the best fit matches very well only the inner and brighter regions of the galaxy for $R \le 10$~arcmin.
As discussed above,  the azimuthally averaged light profiles of NGC~1399 (in both $g$ and $i$ bands) have an extended steeper behaviour for $R\ge R_{break}$ (see right panel of Fig.~\ref{profiles}), which clearly requires a fit with two components.
Therefore, we adopted a Sersic law \citep{Sersic}, given by the equation
\begin{equation}
\label{sersic}
\mu (R) = \mu_e + k(n) \left[ \left( \frac{R}{r_e}\right)^{1/n} -1\right]
\end{equation}

which accounts for the inner and most luminous part of the galaxy, plus an outer exponential function given by
 
\begin{equation}
\mu(R)= \mu_{0} + 1.086 \times R/r_{h}
\end{equation}

In the above formulae  $R$ is the galactocentric distance, $r_e$ and $\mu_e$ are the
effective radius and effective surface brightness respectively, $k(n)=2.17 n - 0.355$ \citep[see][]{Caon1993}, $\mu_{0}$ and $r_{h}$ are the central surface brightness and scale length of the exponential component.

We performed a least-square fit of the azimuthally averaged surface brightness profile in the $g$ band, for $R\ge 0.1$~arcmin, to exclude the power-law core of the galaxy. The structural parameters that give the best fit are the following: $\mu_e=21.5 \pm 0.3$~mag~arcsec$^{-2}$ {  ($\mu_e^B \simeq 22.1$~mag arcsec$^{-2}$)}, $r_e=49.1 \pm 0.7 $~arcsec ($\sim 4.7$~kpc), $n=4.5 \pm 0.7$, $\mu_0=23.4 \pm 0.1$~mag~arcsec$^{-2}$ {  ($\mu_0^B \simeq 24$~mag arcsec$^{-2}$)}, $r_h=292 \pm 4 $~arcsec ($\sim 28$~kpc) and results are shown in the right panel of Fig.~\ref{profiles}.
The total $g$-band magnitude of the exponential component, corrected for the galactic extinction (see Sec.~\ref{result}), is  $m_{tot}^{halo} = 9.13$~mag. {  Thus, compared to the total magnitude of the whole system (see Sec.~\ref{result}), it contributes about 60\% of the total light, where  the ratio is $L_{exp}/L_{gal} =0.60 \pm 0.03$.}

The above parameters are within the range of values derived by \citet{Donzelli2011} for the 205 BCGs having double component profiles. In particular, the scale length of the exponential component measured for NGC~1399 is comparable with the average value of 25~kpc found for this large sample of galaxies. 

\begin{figure}[t]
%\epsscale{1.1}
\includegraphics[scale=0.4]{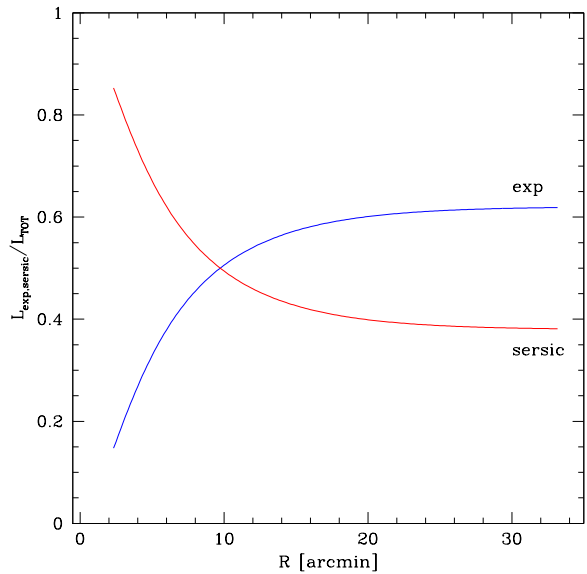}
\caption{\label{BD} Fraction of the Sersic  (red line) and exponential  (blue line) component with respect to the total contribution of both of them as function of the galactocentric radius.}
\end{figure}

{  From the best fit, we derived the total luminosity of the Sersic component and of the exponential one as function of the radius. In Fig.~\ref{BD} are plotted the relative fraction of the two components with respect to the total light profile:} the inner Sersic component dominates the light distribution inside $\sim 1$~arcmin ($\le 0.3 R_e$), where the contribution to the total light is about the 90\%. 
At $R=10$~arcmin, the two components contribute by the 50\% to the total light.
This is consistent with the previous fit performed by \citet{Schombert86}, where a de Vaucouleur's law reproduces very well the light distribution inside 10~arcmin.
The  exponential component in NGC~1399 is the dominant ones for $R > 10$~arcmin, i.e. outside $2R_e$. 

The striking result of this analysis is that the break radius, identified by a "visual inspection" of the light profiles, is a physical scale separating  the regions where the main galaxy and the stellar halo dominate. We had already commented that the transition between the two components occurs at the same radius where  the ellipticity and P.A. of the isophotes become flattened and misaligned (see Fig.~\ref{epsPA}).
Hence one can argue that the bright inner regions are almost spherical and the exponential component is more flattened,  implying a different formation mechanism.

%-------------------------------------------------------------

\section{Discussion: tracing the build-up history of the extended halo at the center of the Fornax cluster}\label{discussion}

We have obtained a new deep mosaic of the Fornax cluster with VST, in the $u$, $g$, $r$ and $i$ bands, which covers an area of about  $3 \times 6$~square degrees around the central galaxy NGC~1399 (see Fig.~\ref{mosaic}). In this paper we focused on the light distribution in the $g$ and $i$ bands inside 2 square degrees around NGC~1399.
The large mosaic obtained with the 1~deg$^2$ field-of-view pointings of OmegaCam and the large integration time allow us to map the surface brightness out to  33~arcmin ($\sim 192$~kpc) from the galaxy center in the $g$ band (see Fig.~\ref{profiles}).
{  At these large distances we are mapping the faint outskirts of NGC~1399, down to $\mu_g = 31.2 \pm 2$~mag arcsec$^{-2}$ {  ($\mu_B \simeq 31.8$~mag arcsec$^{-2}$)}: the stellar envelope, central galaxy plus ICL,  encloses most of the bright galaxies in the core of the Fornax cluster (see Fig.~\ref{1399halo}). 

The main results given by the analysis of the light distribution down to the above levels of surface brightness are the following:
\begin{enumerate}
 
\item we detected a "bridge-like"  stellar stream in the region between NGC~1399 and the nearby early-type galaxy NGC~1387 (see Fig.~\ref{streams}). It is very faint, with $29 \le \mu_g \le 31$~mag~arcsec$^{-2}$ ($29.6 \le \mu_B \le 31.6$~mag arcsec$^{-2}$), and it is about 5~arcmin wide ($\sim 29$~kpc).
Given the very low level of the surface brightness, this  structure was unknown in the previous works;

\item the azimuthally averaged light profile  in the $g$ band is consistent with the presence of two main components:  the bright central part of NGC~1399, which contributes  about  90\% of the light inside 1~arcmin (i.e. $\le 0.3 R_e$) and it is well fitted by a Sersic law with $n=4.5$, and an  exponential outer component from  $R \ge 10$~arcmin, which contributes to about 60\% to the total galaxy light and extends out to about $6 R_e$ (see Fig.~\ref{profiles}). The break radius at $R = 10$~arcmin {  is a physical scale separating} the regions where the two components dominate.

\end{enumerate}

%The 2D light distribution of NGC~1399 at the faint levels of $\mu_g \sim 29 - 31.2$~mag/arcsec$^2$ is not symmetric, appearing  more ''elongated" toward the west side of the cluster.

%In the area presented in this work, we do not detect any other tidal features as evident as those described above. The analysis made by Thorsten et al. (2015, submitted) on the white-filter WFI data covering a similar area of the Fornax cluster, also found that the number of candidate tidal streams is very small even at the surface brightness level of 28 in the V band. 

In this section we discuss the main implications of these results on the build-up history of the center of the Fornax cluster.

 \subsection{What about the origin of the bridge-like stellar stream in the halo region between NGC~1399 and NGC~1387?}\label{stream_formation}
%--- 
Stellar streams of comparable low surface brightness and extension as detected in the outskirts of NGC~1399 are also observed in the halo of several galaxies. Results by  the pilot survey of tidal streams published by \cite{MarDel2010}, as well as those detected around galaxies in the SDSS DR7 archive by \citet{Miskolczi2011} suggest that several morphologies exist  for such a faint stellar structures,  like great circles, spikes, very long tails or also mixed type of streams.  The morphology and stellar population of the stellar streams is strictly related to which kind of gravitational interaction make them and the nature of the galaxies involved in it. 
As discussed by \citet{Duc2015}, the material expelled from a galaxy, which is interacting with a companion massive enough to perturb it, produces a  tidal tail having similar age and metallicity (i.e. colors) to those measured in the parent galaxy. On the other hand, when a low mass galaxy is disrupting in the halo of a massive object it produce a tidal stream, made of material different from that in the massive galaxy.
The system of NGC~2698/99 in the ATLAS$^{3D}$ sample is an example of a tidal interaction producing tails, which also shows a luminous bridge of stars between the two galaxies   \citep[see Fig.~21 in][paper]{Duc2015}.

As mentioned in Sec.~\ref{result}, the stellar bridge-like structure detected between NGC~1399 and NGC~1387 could be the sign of an ongoing interaction between the two galaxies. 
%NGC~1387 is the closest galaxy to NGC~1399, even if NGC~1404 appear closest in projection on the sky, since the systemic velocity differs of only about 123 km/s, while  the velocity difference between NGC~1399 and NGC~1404 is about 522 km/s.
The deep images taken at VST show that NGC~1387 has asymmetric isophotes at  low  surface brightness levels ($29 \lesssim \mu_g \lesssim 31$~mag~arcsec$^{-2}$), which appear more elongated towards the west side. These features are quite evident in the $g-i$ color map, where the galaxy shows a red arc-like structure on this side (see bottom panel of Fig.~\ref{streams}). 
The projected distribution of the blue GCs by \citet{Bassino2006} also shows a bridge between NGC~1387 and NGC~1399, that is, very recently, confirmed by \citet{Dabrusco2016}  on the VST data. These features match very well in space and size the bridge of stellar streams detected in the light distribution.
%These features are tracing a clear contribution to the ICL in the core of the Fornax cluster.

\citet{Bassino2006} claimed that the observed stream of the blue GCs might be caused by a tidal stripping from the outer envelope of NGC~1387 by NGC~1399. This idea was supported by the low number of the blue GCs around NGC~1387 with respect to that of red GCs. 
According to this scenario, the east side of NGC 1387's envelope is stripped away by NGC 1399 and creates the
 stellar bridge observed between the two galaxies.  The red arc on the west side of the galaxy is the surviving  part of that envelope. A hint in favour of this hypothesis comes from the comparison between the average colors measured in the stellar bridge and in the halo of NGC~1387.
We found that the average $g-i$ color in the middle region of the stellar bridge is only $\sim 0.2$~mag bluer than the average colors in the  arc on the west side of NGC~1387, where $g-i = 1.14$~mag (see Fig.~\ref{stream_col}). Taking into account the uncertainties on the color estimates at these level of surface brightness (see Fig.~\ref{color}), the values derived above are all consistent.
Anyway, the color difference between the  surviving red part and the blue stripped part might mean a difference in their origin. Kinematics and numerical simulations will shed light on the interaction between these two galaxies.

 \subsection{Which kind of physical structure is the outer exponential component in NGC~1399?}\label{disk_halo} 
By looking at the light profiles derived for NGC~1399, out to about 192~kpc from the center, it resembles those typically observed for disk galaxies, having a very bright $R^{1/n}$ central component and an outer exponential one  (see Fig.~\ref{profiles}). In this case, NGC~1399 would be a giant S0-like galaxy with a very faint and diffuse face-on disk ($\epsilon \sim 0.2$, see Fig.~\ref{epsPA}), never known in literature.
In order to prove such an hypothesis it is necessary to gather observational and dynamical characteristics for this structure and check if they are consistent with a stellar disk. To do this, we used the photometry presented in this work and the published kinematical data.
By the best fit of the light profile in the $g$ band, the exponential component has a central surface brightness $\mu_0=23.4 \pm 0.1$~mag~arcsec$^{-2}$ ($\mu_0=22.6 \pm 0.1$~mag~arcsec$^{-2}$ in the $r$ band) and a scale length  $r_h=292 \pm 4 $~arcsec ($\sim 28$~kpc).  By comparing these values with the typical ones obtained for a large sample of disk galaxies from SDSS data, in the $\mu_0 - r_h$ plane, they are located along the slope of the constant disk luminosity but in the void region of very low surface brightness galaxies and with a very large disk,   \citep[see Fig.~1 in][and reference therein]{vdKruit}. 
In particular, disk galaxies with a mass comparable to that of NGC~1399, i.e. $\sim10^{12}$~$M_\odot$, have an average scale length of $5.7\pm1.9$~kpc, and the central surface brightness in the $r$ band  is the range  $18 \leq \mu_0 \le 21.8 $~mag~arcsec$^{-2}$. Thus, the "disk" in NGC~1399 would be one magnitude fainter and ten times more extended than any other observed disk in the local universe.

The photometry alone is not enough to exclude the "disk" nature for this component, because we need to prove that such a very faint and enormous disk should be dynamically stable. According to  Toomre's criterion \citep{Toomre1964}, the stability of a galactic disk depends on the stellar velocity dispersion $\sigma$, the epicyclic frequency $k$ and the local surface density $\Sigma$, where the parameter $Q=\sigma k/3.36G\Sigma > 1$.
The most extended stellar  kinematics for NGC~1399 reaches a distance of $R=97$~arcsec ($\sim 9.4$~kpc) \citep{Saglia2000}, thus it does not map the regions of the exponential component. At these large radii, the kinematics is traced by GCs and
planetary nebulae (PNe), out to $\sim 13$~arcmin
\citep{Schuberth2010,McNeil2010}, thus almost out to the region of the
break. The kinematics of red and blue GCs are distinct. The kinematics of the red GCs is consistent with that of the stars in the inner regions of NGC~1399, while blue GCs are more erratic, showing a higher velocity dispersion ($300 \le
\sigma \le 400$~km/s) than the red GCs \citep{Schuberth2008,Schuberth2010}. \citet{Schuberth2008} analyzed the GCs kinematics and their irregularities out to 35~arcmin: they found an increase of the velocity dispersion of the blue GCs beyond 20~arcmin.
The maximum rotation velocity of stars was about 40~km/s inside 9.4~kpc, while those for blue GCs is about 110~km/s at $R\sim 60$~kpc. 
Therefore, the blue GCs are the only traces of the kinematics in the region of the exponential component.
Assuming as average value of the rotation velocity $V_{rot}\sim180$~km/s (corrected for inclination) and as velocity dispersion $\sigma \sim 300$~km/s for the blue GCs, we estimate the $Q$ parameter at the break radius $R=10$~arcmin ($\sim 58$~kpc), where the exponential component starts to dominate the light, and at the last measured point in the surface brightness profiles  $R=33$~arcmin ($\sim 192$~kpc). The local surface density is derived by the surface brightness profile at these radii by using a mass-to-light ratio $M/L=32$~$M_{\odot}/L_{\odot}$ at $R=10$~arcmin and $M/L=42$~$M_{\odot}/L_{\odot}$
 at $R=33$~arcmin. These are derived by using the total mass given by the GCs dynamics \citep[see Fig.19 in][]{Schuberth2008} and the total luminosity integrated at the same radii from the VST $g$ band profile. 
 Given that, we find $Q[R=10]\sim0.85$ and $Q[R=33] \sim 4\times10^{-5}$,  thus $Q < 1$ over the whole extension of the exponential component. This analysis allows us to conclude that such a faint and large exponential component cannot be a dynamically-supported disk.
 Therefore,  we associate the light of inner component to the bright cluster member NGC~1399 and the extended outer component to its stellar halo. The system {\it galaxy plus halo} has a total magnitude in the $g$ band of 8.57~mag (see Sec.~\ref{result}), therefore it has total luminosity $L \sim 2 \times 10^{11}$~L$_{\odot}$. According to the M/L ratio at 192~kpc estimated above, we found that this system has a total mass M~$\sim 7 \times 10^{12}$~M$_{\odot}$.}

 \subsection{What is the vision from other dynamical tracers in the halo region?}\label{other_tracers}
\begin{figure}[t]
%\epsscale{1.1}
\includegraphics[scale=0.4]{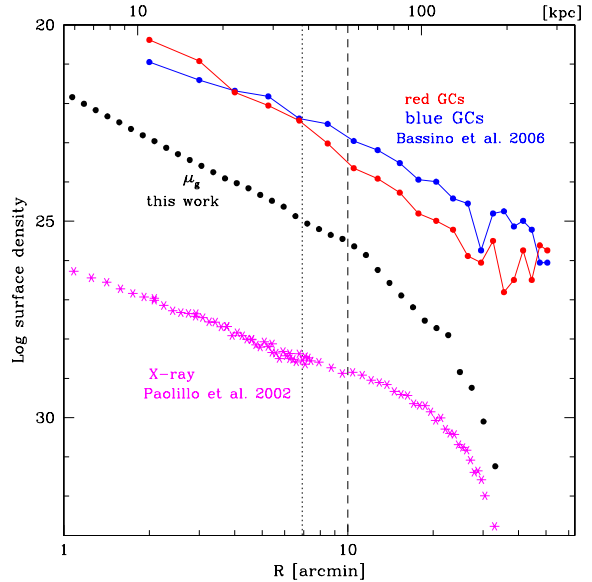}
\caption{\label{tracers} Overview of the surface density profiles of the blue and red GCs (blue and red circles), by \citet{Bassino2006}, and X-ray (magenta asterisks), by \citet{Paolillo2002} compared with the surface brightness profile in the $g$ band (black circles), as function of the galactocentric radius, for NGC~1399. The surface density profiles are arbitrarily shift in the ordinate scale. The dotted line indicates the "transition" region where the blue GCs and the X-ray density profile show a different slope with respect to the inner regions. The dashed line indicate the distance of 10~arcmin where the contribution of the inner component and stellar halo the light distribution is of about the 50\% (see Sec.~\ref{fit} for details).}
\end{figure}

The large-scale study of the GCs in NGC~1399 published by \citet{Dirsch2003} and \citet{Bassino2006} provide the color and spatial distribution up to the projected distance of 52~arcmin. They found that the blue GCs dominate  in the outer regions and  their radial distribution show a slope change at $R \sim  7$~arcmin, while the red GCs have a single power law profile on the entire range of radii (see Fig.~\ref{tracers}). 
The radial distribution of the red GCs follows the slope of the  surface brightness out to the break radius, while the blue GCs do not seem to match the steep light profile for $R \ge R_{break}$, being shallower. 

The X-ray halo of NGC~1399 extends out to about 90 kpc and it is very
asymmetric; \citet{Paolillo2002} noted the presence of several
components: the 'galactic' halo extends out to $\sim 7$ arcmin and is
offset toward the west with respect to the optical galaxy centroid,
while on larger scales the 'cluster' halo is elongated in the opposite
direction. The transition between the 'galactic' and 'cluster' X-ray
density profile occurs at $R \sim 7$~arcmin, which is consistent with the
value where the blue GCs radial distribution starts to dominate over
the red GC component, and it is a few arc minutes smaller than the break
radius identified in the light distribution (see Fig.~\ref{tracers}).  

%The correlation between the asymmetries in the light  and X-Ray distributions has also been observed in NGC~3311, in the core of the Hydra cluster, where the extended X-Ray emission NE of NGC~3311 is supposed to be a morphology reminiscent of the off-centered halo in the stellar light distribution \citep{Arnaboldi2012}. For NGC~1399 we have checked that the light distribution of the stellar halo has on average the same light profile at different position angles, and that some deviations, which appear as an excess of light, are detected at larger radii on West side (see Fig.~\ref{axis}).

As a conclusion of the cross-analysis with different tracers we find 
that  there is a "transition" region (very close in radius)
where the light distribution, the blue GCs and the X-ray density
profile show a different slope with respect to the inner regions (see Fig.~\ref{tracers}).
In correspondence of this region the kinematics of
blue GCs is different from that of the red GCs, which is consistent
with that of the stars in the inner regions, while the analysis of the
X-ray emitting gas supports the presence of different dynamical
components. Moreover, as pointed out in Sec.~\ref{result}, the 2D density
distributions of stars, blue GCs and X-ray consistently show an
asymmetry on the west side, toward the nearby early-type galaxy NGC~1387.

The above results confirm those derived by \citet{Hilker1999} that adopted a similar approach. By comparing the light distribution in NGC~1399, down to about 24 mag~arcsec$^{-2}$ in the V band, with the surface density of  GCs and  X-ray,  and with the surface brightness of dE/dS0 galaxies, they found that the profile slopes of the blue GCs population, X-ray and the cD halo light are very similar in extension and slope. As suggested by \citet{Hilker1999}, the agreement in the above properties might suggest that these components share a common history.

 \subsection{What do theoretical studies on the halo formation predict for the shape of the light profile?}\label{halo_shape}
According to \citet{Cooper2013}, the {\it break radius} in the total
stellar mass density sets the transition region in the galaxy halo
between the stars formed in situ to those captured from accretion
events. The shape of the light profile for $R \ge R_{break}$ could be
indicative of which kind of accretion mechanism shaped the galaxy halo in the
past. As discussed by \citet{Deason2013}, when plotted in logarithmic
scale, some outer light profiles are steeper, named as ''broken", in contrast to 
 others that show an ''excess of light" at large radii. 
The broken profiles are generally associated to short and
more recent accretion, possibly unmixed as they trace the orbit of the
accretion, while the shallower profiles could result from minor merging
events taking place over long timescales, showing a larger degree of mixing.
To produce such a kind of broken profile, simulations found that the  accreted material from satellites has similar apocenters or, alternatively, it results from the accretion of a massive satellite \citep{Deason2013}. 
As investigated by \citet{Deason2013}, the Milky Way also shows a broken profile of its stellar halo, and a break is detected at about 20-30 kpc from the center. On the other hand, no break is detected in the halo profile of M31. The kinematics can discriminate between the two cases, since if a massive satellite was accreted, the metal-rich population of the accreted stars shows a peak in the  velocity dispersion at $R \ge R_{break}$.

In the case of NGC~1399, the light profile deviates from a smooth Sersic profile, with $n\simeq4$, for $R_{break} \ge 10$~arcmin, showing a
steeper exponential profile with a scale length $r_h \simeq 4.8$~arcmin ($\sim 28$~kpc, see right panel of Fig.~\ref{profiles}).  
{  Similar to NGC~1399, also the BCGs studied by \citet{Seigar2007} have a stellar halo with an exponential light profile.}
Differently from NGC~1399, the light profile of M87 is shallower with increasing radius and it was reasonably well-fitted by either a single Sersic law or a double $r^{1/4}$ \citep{Jan2010}. As found from \citet{Bender2015}, also the light distribution of the cD galaxy NGC~6166 in Abell~2199 is well fitted by a single Sersic law with a high value of the exponent ($n \sim 8$), therefore it appears shallower towards outer regions. According to the theoretical predictions, the differences in the halo profiles can reflect a different assembly history of this component in the BCGs. 
In NGC~1399, as  discussed above, the regions after the break radius are purely explored with any kinematical tracers. The only information is close to the break radius and indicates that a peak in the velocity dispersion is observed in the metal-poor blue GCs \citep{Schuberth2008}, but this is not enough to draw any conclusion on this point. 

Therefore, based only on the new deep photometry, we might address some points useful for the interpretation in the context of the stellar halo formation scenarios, which are not   conclusive and need to be confirmed by a kinematical analysis.

First, one should take into account that the exponential stellar halo in NGC~1399 is the dominant  component almost everywhere outside the $2 R_e$ and it contributes to about the 60\% of the total light. Thus, it might be reasonably to think that the material forming this component could be stripped by a massive satellite, during a close passage around NGC~1399.  
Previous observational studies on NGC~1399 strongly suggests that
the core of the Fornax cluster  underwent  merging events, which led to the
accretion of the majority of blue GCs \citep{Schuberth2010}. 
If any kind of  gravitational interaction happened in a recent epoch, we should found signs of it, in the form of tidal tails or streams, in the light distribution. Therefore, the frequency and the spatial distributions of such a features are strong constraints on the timescale and orbits of the accretion event, respectively.

{  Differently from the Virgo cluster, the deep images of the core of the Fornax cluster  do not reveal bright and extended structures of this type, except for the red tail detected at $R\sim 3$~arcmin and the faint streams towards NGC~1387 (see Sec.~\ref{result}, Fig.~\ref{streams}). This remains valid 
even considering that the Virgo cluster is about 3~Mpc closer,  thus the faint structure in the light distribution detected in the Fornax cluster would have an absolute magnitude of only $\sim 0.3$~mag brighter.}
On the other hand, at the same faint level of surface brightness, around the giant bright galaxies M49, M86 and M87  several tidal tails are detected, some of them very extended, up to 100~kpc \citep{Mihos2005,Rudick2010,Jan2010,Longobardi2015,Cap2015}. 
This observational fact might suggest that we are tracing two different epochs of the halo assembly in the two clusters: the Fornax cluster could be in a more dynamically evolved phase \citep{Grillmair1994, Jordan2007}, where most of the gravitational interactions between galaxies have already taken place and streams and tails are already dispersed into the intracluster medium or/and are below the observable surface brightness level.  

{  
 \subsection{Stellar halo mass fraction in NGC~1399: observations versus theoretical predictions}\label{halo_ratio} 
By fitting the light profile,  we found that exists a physical break radius in the total light distribution at $R=10$~arcmin ($\sim58$~kpc) that sets the transition region between the bright central galaxy and the outer exponential halo, and that the stellar halo contributes for  60\% of the total light of the galaxy (Sec.~\ref{fit}).  

In order to estimate the stellar mass fraction of the halo for NGC~1399, we need to account for the stellar mass-to-light ratio  value for both the central galaxy and the halo. The color profiles shown in Fig.~\ref{color} suggest that the stellar halo (for $R\ge 10$~arcmin) has $g-r$ color similar to those measured in the central galaxy, thus we can assume a similar stellar M/L for both components. According to an average $g-r \sim 0.8$, from the stellar population synthesis model \citep{Vazdekis2012,Ricciardelli2012}, using a Kroupa IMF,  M/L$\sim4$.
Therefore, the total stellar mass of the whole system, galaxy plus halo, is M$_{tot} \simeq 6.6 \times 10^{11}$~M$_{\odot}$, while the stellar mass in the halo is M$_{halo} \simeq 4 \times 10^{11}$~M$_{\odot}$. The stellar mass fraction is the same we found for the light, which is $0.58\pm0.03$. 
As shown in Fig.~\ref{halo_ratio}, this value is comparable with those derived for other BCGs with total stellar mass similar to NGC~1399, which varies from 0.47 to 0.9 for the galaxies in the sample studied by \citet{Seigar2007}, and it is  about 0.65 for the bright BCG NGC6166 \citep{Bender2015}. 

The observed halo mass fraction for NGC~1399 and for other published data are compared with the predictions from numerical simulations by \citet{Cooper2013}, for galaxies with a dark halo virial mass within the range $13 < \log M_{200}/M_{\odot} < 14$, and from the {\it Illustris}  hydrodynamic cosmological simulations by \citet{Rodriguez2015}. 
According to the dynamical models performed by \citet{Schuberth2010}, the virial mass for NGC~1399 is in the range $5 \times 10^{12} - 7 \times 10^{13} M_{\odot}$.
On average, taking into account the uncertainties in M/L estimate, the stellar mass fraction derived for the halo of NGC~1399 fully agrees with those predicted by the Illustris simulations, while it is at the lower limits of those by \citet{Cooper2013}. 
The stellar mass fractions of the halos in the less luminous BCGs studied by \citet{Seigar2007} are more consistent with simulations, except for NGC~4874, where deviations could be due to a wrong estimate of the M/L ratio and/or of the halo stellar mass fraction.
On the other hand, the brightest galaxy of the selected sample, NGC~6166, is far from the theoretical predictions: this could be due to an unexplored regions in the simulations, since Fig.12 in \citet{Cooper2013} shows that there are no values for an halo stellar mass larger than $10^{12}$~M$_{\odot}$.
We noticed that the region of the stellar halo mass fraction larger 0.8 seems devoid of observed stellar halos. The same effect seems to be also present in the same diagram obtained by \citet{Trujillo2015} (see Fig.12 of that paper), where the stellar mass fractions derived for less massive galaxies (in the range $12 < \log M_{200}/M_{\odot} < 12.5$) occupy the bottom half part of the region given by theoretical predictions.
In both cases, this could be to an incomplete sample. Anyway, this aspect needs to be further investigated on both observational and theoretical side.

Assuming that the stellar halo has been formed by accretion then we find consistency with simulations. This means that  60\% of the total light we measured in the core of the Fornax cluster for $R \le 192$~kpc is formed via multiple accretion events. According to \citet{Cooper2013}, the stellar mass of accreted satellites for halos of the most massive galaxies (i.e. $13 < \log M_{200}/M_{\odot} < 14$) varies in the range $10^{8} - 10^{11}$~M$_{\odot}$. While the lower limit of this interval is consistent with the typical mass for  ultra compact dwarf galaxies \citep{Chilingarian2011}, the upper limit is  that for a brighter and larger object. This might suggest that the halo of NGC~1399 has also gone through a major merging event. Given the absence of any tidal tail in the deep VST images, the epoch of this strong interaction goes back to an early formation epoch. 
For comparison, the total stellar mass of NGC~1387, which is interacting with NGC~1399 (see Sec.~\ref{stream_formation}),  is about $8 \times 10^{10}$~M$_{\odot}$, assuming a M/L similar to that derived for NGC~1399, or even higher considering that NGC~1387 has on average redder colors than NGC~1399 (Iodice et al. in preparations). 

The {\it Illustris}  hydrodynamic cosmological simulations by \citet{Rodriguez2015} also predict a {\it transition radius} at which the halo component starts to dominate over the stars formed in-situ. Such a radius is a function of the total mass of the galaxy, ranging from $\sim 4-5$ effective radii for medium-sized galaxies ($M_{star} \sim 10^{10} - 10^{11}$~M$_{\odot}$) to a fraction of the effective radius for very massive galaxies ($M_{star} \sim 10^{12}$~M$_{\odot}$). 
For NGC~1399, the total stellar mass is M$_{tot} \simeq 6.6 \times 10^{11}$~M$_{\odot}$ and the break radius in term of effective radius of the light distribution is $R_{break} = 1.7 R_e$ (see Sec.~\ref{ellipse}). Such a value is comparable with the predictions from {\it Illustris} simulations on the break radius, which varies from $\sim 0.2 R_{half}$ for the most massive galaxies (stellar masses $\sim 10^{12}$~M$_{\odot}$) to $\sim  3.5 R_{half}$ for galaxies with stellar mass of $\sim 10^{11}$~M$_{\odot}$.}

\begin{figure}[t]
%\epsscale{1.1}
\includegraphics[scale=0.4]{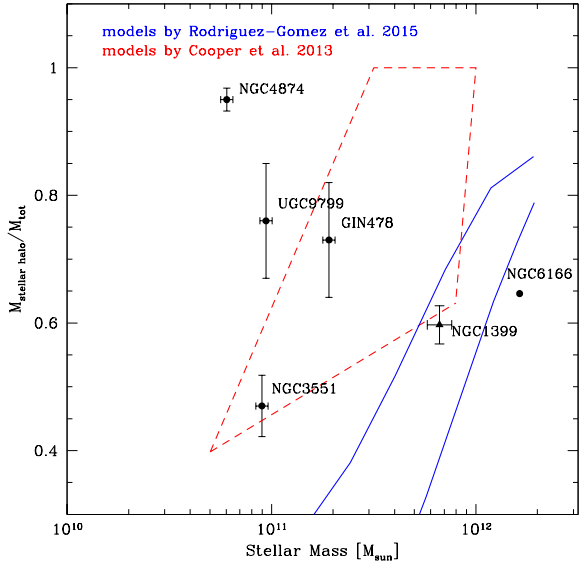}
\caption{\label{halo_ratio} Stellar halo mass fraction versus total stellar mass of the galaxy. Triangle indicates  the value derived for NGC~1399, circles are for other BCGs from literature \citep{Seigar2007,Bender2015}. The dashed red region indicates the average predictions from theoretical models by \citet{Cooper2013}. The blue region indicates the average predictions from the Illustris hydrodynamic simulations  by \citet{Rodriguez2015}.}
\end{figure}

 \subsection{Halo vs ICL}\label{icl} 
The presence of ICL in nearby clusters was established by several works in the recent years \citep[see][as review]{Zibetti2008}. As mentioned in the Sec.~\ref{intro}, the extended stellar halo in BCGs and the intracluster stellar component overlap and, therefore, it is hard to separate each  contribution by deep photometry alone. Surface photometry indicates that the ICL surface brightness is detected in the range between $\sim26$ to $\sim32$~ mag~arcsec$^{-2}$ in $r$ band, and it declines more steeply than the light profile of the galaxies in the cluster. The contribution of the ICL to the total light of most clusters is about $30\%$  \citep{Gonzalez2005,Zibetti2005}.
On average, this component is more elongated than the BCG and the major axes of BGCs and the ICL are 
 almost aligned  ($\Delta P.A. \le 10$~degrees), even if higher degrees of misalignments are also measured in some cases.
The deep imaging of the Virgo cluster shows that other than the numerous thin ICL streams,  part of this component is in a more diffuse form, locked in the extended halo around M87, traced up to 150~kpc \citep{Mihos2015}.

{  For NGC~1399,  the deep VST photometry allows us to map the  surface brightness levels typical for the ICL contribution.
The azimuthally averaged surface brightness profile derived in this work shows a steeper exponential decline for $ \mu_g \ge 26.5$~mag~arcsec$^{-2}$ (i.e. $ \mu_r \ge 25.65$~mag~arcsec$^{-2}$ since $g-r \sim0.85$) and it reaches a $ \mu_g =31.2$~mag~arcsec$^{-2}$ at $\sim192$~kpc from the analyz (see Fig.~\ref{profiles}).  As discussed in detail in the previous sections, this component is the stellar halo around NGC~1399, which is very extended and contributes to about 60\% of the total light.
We may suppose that part of this huge amount of light comes from the contribution of the ICL, which could vary from 10\% to 30\% as  was estimated in other cluster of galaxies  \citep{Gonzalez2005,Zibetti2005}.

A hint for extra-light is detected from the light profiles extracted at several P.A.s, in the regions between galaxies on the west side (see Sec.~\ref{result} and Fig.~\ref{axis}), in the range of surface brightnesses  $27 \le \mu_g \le 28$~mag~arcsec$^{-2}$ and for $15 \le R \le 20$~arcmin.
At fainter level of surface brightnesses, the stellar bridge between NGC~1399 and NGC~1387 is about $1-1.5$~mag~arcsec$^{-2}$ brighter than the average value of the diffuse component ($\mu_g \sim 31$~mag~arcsec$^{-2}$, see Fig.~\ref{stream_col}).
As found in other BCGs \citep{Gonzalez2005,Zibetti2005}, the ellipticity and P.A. profiles of the stellar envelope show that it  is flatter with respect to the inner galaxy light and both are misaligned by  about 10 degrees (see Fig.~\ref{epsPA}).

However, photometry alone does not allow us to separate the gravitationally bound part of NGC 1399's very extended stellar halo from that of the intracluster light. There is a hint from the kinematics of the GCs for a higher velocity dispersion of the blue population at large radii of about 20~arcmin \citep{Schuberth2010}, but this point needs further confirmation with new data.  Therefore,
at this stage, no definitive conclusion can be drawn and kinematical studies are crucial to disentangle the bound and unbound components in the stellar halo of NGC~1399. }

%------------------

{  \section{Summary}

In conclusion, the most important and first results of the Fornax Deep Survey with VST can be summarised as follows:
\begin{enumerate}
\item deep photometry allowed us to detect the faint stellar counterpart of the blue GCs stream in the intracluster region on the west side of NGC~1399 and towards NGC~1387. By analyzing the integrated colors of this feature, we claim that it could be  explained by the ongoing interaction between the two galaxies, where the outer envelope of NGC~1387 on its east side is stripped away;

\item in the core of the Fornax cluster a very extended envelope surrounding the luminous galaxy NGC~1399 is present. This component, which is well fitted by an exponential law, comprises  60\% of the total integrated light out to 192~kpc, included the ICL fraction;

\item The surface brightness profile of NGC~1399 shows a physical break radius in the total light distribution at $R=10$~arcmin ($\sim58$~kpc), i.e. $1.7 R_e$, that sets the transition region between the bright central galaxy and the stellar halo. Such a value is comparable with the predictions from {\it Illustris} simulations on the break radius for galaxies in the range of stellar masses $10^{11} - 10^{12}$~M$_{\odot}$;

\item  comparing with the numerical simulations of  stellar halo formation for the most massive BCGs (i.e. $13 < \log M_{200}/M_{\odot} < 14$), we found that the observed stellar halo mass fraction is consistent with a halo formed through the multiple accretion of progenitors having stellar masses in the range $10^{8} - 10^{11}$~M$_{\odot}$. This might suggest that the halo of NGC~1399 has also went through a major merging event. 

\item The absence of a significant number of luminous stellar streams and tidal tails between the galaxies in the core of the Fornax cluster suggests that the epoch of this strong interaction goes back to an early formation epoch. Thus, the extended stellar halo around NGC~1399 is characterised by a more diffuse  and  well-mixed  component, including the ICL, with respect to that of M87 or M49 in the Virgo cluster \citep{Mihos2005,Cap2015}. This is an indication that the Fornax cluster could be in a more dynamically evolved phase than the Virgo cluster.
\end{enumerate}}

%-------------------------------------------------------------------------


%% Included in this acknowledgments section are examples of the
%% AASTeX hypertext markup commands. Use \url without the optional [HREF]
%% argument when you want to print the url directly in the text. Otherwise,
%% use either \url or \anchor, with the HREF as the first argument and the
%% text to be printed in the second.

 \acknowledgments{ This work is based on visitor mode observations taken at the ESO La Silla Paranal Observatory within the VST GTO Program ID 094.B-0496(A). 
{  The authors wish to thank the anonymous referee for his/her comments and suggestions
    that allowed us to greatly improve  the paper. }
 Authors wish to tank ESO for the financial contribution given for the visitor mode runs at the ESO La Silla Paranal Observatory. Enrichetta Iodice wish to thank the ESO staff of the Paranal Observatory for the support during the observations at VST. E.I. is also very grateful to T. de Zeeuw, M. Arnaboldi, O. Gerhard and I. Truijllo for the discussions and suggestions on the present work. Michele Cantiello acknowledges financial support from the project EXCALIBURS (PRIN INAF 2014, PI G. Clementini). Nicola R. Napolitano and M.Paolillo acknowledges the support of PRIN INAF 2014 "Fornax Cluster Imaging and Spectroscopic Deep Survey". }

\end{document}